\documentclass[%
reprint,
superscriptaddress,
frontmatterverbose,
showpacs,
preprintnumbers,
nofootinbib,
nobibnotes,
amsmath,amssymb,
aps,
prc,
]{revtex4-2}
\usepackage[pdftex]{graphicx}
\usepackage[caption=false]{subfig}
\usepackage{amsmath}
\usepackage{mathtools}
\usepackage{dcolumn}
\usepackage{bm}
\usepackage[pdftex]{color}
\usepackage{CJK}
\usepackage[T1]{fontenc}
\usepackage{float}
\usepackage{mathrsfs}
\def\ket#1{\left|{#1}\right\rangle}
\def\braket#1#2{\left\langle{#1}\middle|{#2}\right\rangle}
\def\brakket#1#2#3{\left\langle{#1}\middle|{#2}\middle|{#3}\right\rangle}

\def\nuc#1#2#3{{}^{#2}_{#3}\mathrm{#1}}
\def\urm#1{\scriptstyle{\text{\textrm{\textmd{\textup{#1}}}}}}
\def\uurm#1{\scriptscriptstyle{\text{\textrm{\textmd{\textup{#1}}}}}}

\let\temp\epsilon
\let\epsilon\varepsilon
\let\varepsilon\temp
\let\temp\relax
\let\temp\phi
\let\phi\varphi
\let\varphi\temp
\let\temp\relax

\DeclareMathOperator{\softplus}{softplus}

\begin{document}
%
\begin{CJK*}{UTF8}{}
  \preprint{RIKEN-iTHEMS-Report-24}
  \title{A deep neural network approach to solve the Dirac equation}
  \author{Chuanxin Wang (\CJKfamily{gbsn}{王传新})}
  \affiliation{
    College of Physics,
    Jilin University,
    Changchun 130012, China}
  \affiliation{
    RIKEN Center for Interdisciplinary Theoretical and Mathematical Sciences (iTHEMS),
    Wako 351-0198, Japan}
  \author{Tomoya Naito (\CJKfamily{min}{内藤智也})}
  \email{
    tnaito@ribf.riken.jp}
  \affiliation{
    RIKEN Center for Interdisciplinary Theoretical and Mathematical Sciences (iTHEMS),
    Wako 351-0198, Japan}
  \affiliation{
    Department of Physics, Graduate School of Science, The University of Tokyo,
    Tokyo 113-0033, Japan}
  \author{Jian Li (\CJKfamily{gbsn}{李剑})}
  \email{
    jianli@jlu.edu.cn}
  \affiliation{
    College of Physics, Jilin University,
    Changchun 130012, China}
  \author{Haozhao Liang (\CJKfamily{gbsn}{梁豪兆})}
  \email{
    haozhao.liang@phys.s.u-tokyo.ac.jp}
  \affiliation{
    Department of Physics, Graduate School of Science, The University of Tokyo,
    Tokyo 113-0033, Japan}
  \affiliation{
    RIKEN Center for Interdisciplinary Theoretical and Mathematical Sciences (iTHEMS),
    Wako 351-0198, Japan}
  \date{\today}
  \begin{abstract}
    We extend the method from [Naito, Naito, and Hashimoto, Phys.~Rev.~Research \textbf{5}, 033189 (2023)] to solve the Dirac equation not only for the ground state but also for low-lying excited states using a deep neural network and the unsupervised machine learning technique. 
    The variational method fails because of the Dirac sea, which is avoided by introducing the inverse Hamiltonian method.
    For low-lying excited states, two methods are proposed, which have different performances and advantages.
    The validity of this method is verified by the calculations with the Coulomb and Woods-Saxon potentials.
  \end{abstract}
  \maketitle
\end{CJK*}
%
\section{Introduction}
\label{sec:level1}
\par
Nowadays, machine learning is being utilized in various fields of physics~\cite{
  carleo2019machine}.
Among machine learning algorithms, the unsupervised machine learning is a powerful technique to optimize neural networks without training datasets.
With the powerful optimization ability, it is an ideal tool to solve variational problems.
\par
The key idea of applying the unsupervised machine learning to quantum mechanics is to find the loss function suitable for the system.
By using the energy as the loss function, searching the minimum of the loss function is equivalent to finding the ground state of the system.
\par
The pioneer work solved spin systems by using the Boltzmann machine~\cite{
  doi:10.1126/science.aag2302}.
Subsequent research enhanced this approach, achieving higher accuracy through the introduction of a pair-wise ansatz~\cite{
  PhysRevB.96.205152} and deterministic time evolution~\cite{
  Carleo2018}.
The extension to the excited-state calculation with two different ansatz were proposed in Refs.~\cite{
  PhysRevLett.121.167204,
  doi:10.7566/JPSJ.89.054706}.
Other approaches to solve spin systems included to apply Jastrow-Slater wave function ansatz~\cite{
  PhysRevB.102.205122}
and hidden-fermion one~\cite{
  robledo2022fermionic}.
Reference~\cite{
  yoshino2023spatially}
proposed a generalization model for spins, which was to take perceptrons as the transfer matrix of a spin-glass Hamiltonian.
\par
For continuous systems, the design of the machine learning algorithm should consider the (anti)symmetrization of the system.
Several models of bosonic systems were solved with a fully connected deep neural network (DNN) in Ref.~\cite{
  doi:10.7566/JPSJ.87.074002}. 
Reference~\cite{
  PhysRevResearch.4.023138}
employed the Deep Sets neural-network architecture to solve periodic systems.
For fermionic systems, from electrons to nucleons, the Jastrow-Slater ansatz was a successful application to consider the antisymmetrization~\cite{
  pfau2020ab,
  cassella2023discovering,
  lou2024neural,
  ruggeri2018nonlinear,
  hermann2020deep,
  li2022ab,
  PhysRevLett.127.022502,
  gnech2022nuclei,
  yang2022consistent,
  PhysRevC.107.034320,
  PhysRevResearch.5.033062,
  fore2024investigating,
  PhysRevResearch.4.043178,
  wilson2023neural}.
Recent progress on this direction is summarized in Ref.~\cite{
  Hermann2023Nat.Rev.Chem.7_692}.
Another application was based on the Jordan-Wigner mapping or similar mapping methods from the fermionic system into a spin system
to use the Boltzmann machine~\cite{
  choo2020fermionic}.
Reference~\cite{
  PhysRevResearch.5.033189}
proposed a DNN model to solve the ground state and excited states in the coordinate space, which included symmetrization for bosons or antisymmetrization for fermions.
Soon this method was extended with spin and isospin degrees of freedom by applying a non-fully connected DNN structure~\cite{
  wang2024neural}.
However, all these works focused on the non-relativistic case except for Ref.~\cite{LORIN2022108474}, which employed physics-informed neural networks to solve the time-dependent Dirac equation.
The relativistic formulation using the Dirac equation is sometimes crucial to describe electronic structure of atoms~\cite{
  armstrong1966relativistic},
molecules~\cite{
  ernzerhof2010conjugated},
and nuclear systems~\cite{
  serot2004building}.
Notably, the unsupervised machine learning has not been applied to the Dirac equation yet;
a fundamental challenge in variational treatments of the Dirac equation is variational collapse A fundamental challenge in variational treatments of the Dirac equation is variational collapse.
Because the Dirac sea exists, the ground-state energy does not correspond to the lowest eigenvalue of the Dirac Hamiltonian.
Direct energy minimization therefore causes the loss function to plunge into the Dirac sea---a phenomenon known as variational collapse.
\par
In this paper, we propose an unsupervised DNN method to solve the Dirac equation, which is an extension of the method proposed in Ref.~\cite{
  PhysRevResearch.5.033189}.
We apply the inverse Hamiltonian method~\cite{
  PhysRevC.82.057301}
to avoid the variational collapse.
To calculate low-lying excited states, we propose two methods: One is adjusting a parameter contained in the inverse Hamiltonian method, and the other is an extension based on the method in Ref.~\cite{
  PhysRevResearch.5.033189},
which applies the orthonormal condition to eliminate components of the lower excited states and the ground state in the DNN output.
Throughout our test, both two methods are capable of generating precise energies and wave functions, while the inverse Hamiltonian method is easier to use and the orthonormal method converges faster.
\par
This paper is organized as follows: 
In Sec.~\ref{Sect:II}, we discuss the Dirac equation with spherical symmetry, and introduce the inverse Hamiltonian method to solve the ground state, while the excited states are solved by both the inverse Hamiltonian and the orthonormal method.
In Sec.~\ref{Sect:III}, we introduce our DNN model.
The results of Coulomb and Wood-Saxon potential are shown in Secs.~\ref{Sect:IV} and~\ref{Sect:V}, respectively.
Section~\ref{Sect:VI} is the summary of our work.
%
\section{Theoretical framework}
\label{Sect:II}
%
%
\subsection{The Dirac equation with spherical symmetry}
\par
The Dirac equation reads~\cite{
  grant2007relativistic}
\begin{equation}
  \label{eq1}
  H_{\urm{D}}
  \varphi
  =
  E
  \varphi,
\end{equation}
where an eigenvalue $ E $ includes the rest mass $ mc^2 $. 
We define the internal energy $ \epsilon $ by
\begin{equation}
  \label{eq2}
  \epsilon = E - mc^2.
\end{equation}
The one-body Dirac Hamiltonian $ H_{\urm{D}} $ in the coordinate space reads
\begin{equation}
  H_{\urm{D}} 
  = 
  c \bm{\alpha} \cdot \bm{p} 
  + 
  \beta \left( mc^2 + S \left( \bm{r} \right) \right) 
  + 
  V \left( \bm{r} \right),
\end{equation}
where $ \bm{p} $ denotes the momentum operator,
and $ S \left( \bm{r} \right) $ and $ V \left( \bm{r} \right) $ denote the scalar and vector potentials, respectively.
The $ \bm{\alpha} $ and $ \beta $ are the Dirac matrices, which are
\begin{subequations}
  \begin{align}
    \alpha_i
    & =
      \begin{pmatrix}
        0        & \sigma_i \\
        \sigma_i & 0
      \end{pmatrix}
                   \qquad \text{($ i = x $, $ y $, $ z $)}, \\
    \beta
    & = 
      \begin{pmatrix}
        I_2 & 0    \\
        0   & -I_2
      \end{pmatrix}
  \end{align}
\end{subequations}
with the Pauli matrices $ \sigma_i $ and the $ 2 \times 2 $ identity matrix $ I_2 $.
We define $ H'_{\urm{D}} $ by
\begin{align}
  H'_{\urm{D}} 
  & =
    H_{\urm{D}} - mc^2 \notag \\
  & =
    \bm{\alpha} \cdot \bm{p} 
    + 
    \beta \left( mc^2 + S \left( \bm{r} \right) \right) 
    + 
    V \left( \bm{r} \right)
    -
    mc^2,
  \label{eq4}
\end{align}
where the eigenvalue of $ H'_{\urm{D}} $ is the internal energy $ \epsilon $.
\par
We assume that the potentials are spherically symmetric, i.e., $ V \left( \bm{r} \right) = V \left( r \right) $ and $ S \left( \bm{r} \right) = S \left( r \right) $.
Then, eigenfunctions of $ H'_{\urm{D}} $ can be written as 
\begin{subequations}
  \begin{align}
    \varphi_{n \kappa j_z} \left( \bm{r} \right)  
    & =
      \frac{1}{r}
      \begin{pmatrix}
        F_{n \kappa} \left( r \right)   \mathcal{Y}_{j j_z}^{A} \left( \theta, \psi \right) \\
        iG_{n \kappa} \left( r \right)  \mathcal{Y}_{j j_z}^{B} \left( \theta, \psi \right)
      \end{pmatrix}
    \qquad
    \text{($ \kappa = - j - 1/2 $)}, 
    \label{eq6a} \\
    \varphi_{n \kappa j_z} \left( \bm{r} \right)  
    & =
      \frac{1}{r}
      \begin{pmatrix}
        F_{n \kappa} \left( r \right)  \mathcal{Y}_{j j_z}^{B} \left( \theta, \psi \right) \\
        iG_{n \kappa} \left( r \right) \mathcal{Y}_{j j_z}^{A} \left( \theta, \psi \right)
      \end{pmatrix}
    \qquad
    \text{($ \kappa = j + 1/2 $)}, 
    \label{eq6b}
  \end{align}
\end{subequations}
where $ F_{n \kappa} \left( r \right)  $ and $ G_{n \kappa} \left( r \right) $ denote the large and small components of a radial wave function, respectively.
The subscripts $ n $ and $ j_z $, respectively, denote the principal quantum number and the $ z $-projection of the total angular momentum $ j $.
For a given $ j $,
$\kappa = j + 1/2 $ if the orbital angular momentum number $ l $ satisfies $ l = j + 1/2 $,
and $ \kappa = - j - 1/2 $ if $ l $ satisfies $ l = j - 1/2 $.
The detailed forms of the tensor spherical harmonics are given by
\begin{subequations}
  \begin{align}
    \mathcal{Y}_{j j_z}^{A}  \left( \theta, \psi \right)
    & =
      \frac{1}{\sqrt{2l + 1}}
      \begin{pmatrix}
        \sqrt{l + m + 1}
        Y_{l, l_z}   \left( \theta, \psi \right)   \\
        \sqrt{l - m} 
        Y_{l, l_z+1} \left( \theta, \psi \right)
      \end{pmatrix}, \\
    \mathcal{Y}_{j j_z}^{B}  \left( \theta, \psi \right)
    &=
      \frac{1}{\sqrt{2l + 3}}
      \begin{pmatrix}
        -\sqrt{l - m + 1}
        Y_{l + 1, l_z}   \left( \theta, \psi \right) \\
        \sqrt{l + m + 1} 
        Y_{l + 1, l_z+1} \left( \theta, \psi \right)
      \end{pmatrix},
  \end{align}
\end{subequations}
respectively, where $ Y_{l, l_z}  \left( \theta, \psi \right)$ is a spherical harmonics with $ l_z $ denoting the $ z $-projection of $ l $.
\par
After the separation of radial and angular parts, the radial Dirac equation is
\begin{equation}
  \label{eq8}
  H'_{\urm{D} r}
  \phi_{n \kappa} \left( r \right)
  =
  \epsilon_{n \kappa}
  \phi_{n \kappa} \left( r \right),
\end{equation}
where $ H'_{\urm{D} r} $ and $ \phi_{n \kappa} $, respectively, read
\begin{align}
  H'_{\urm{D} r}
  & =
    \begin{pmatrix}
      V \left( r \right)  + S \left( r \right)         & - \frac{d}{dr} + \frac{\kappa}{r}                \\
      \frac{d}{dr} + \frac{\kappa}{r}                  & V \left( r \right)  - S \left( r \right) - 2mc^2 
    \end{pmatrix},
                                                         \label{eq7} \\  
  \phi_{n \kappa} \left( r \right)
  & = 
    \begin{pmatrix}
      F_{n \kappa} \left( r \right) \\
      G_{n \kappa} \left( r \right) 
    \end{pmatrix}.
\end{align}
According to Eq.~\eqref{eq8}, $G_{n \kappa} \left( r \right)$ can be calculated by
\begin{equation}
  \label{eq9}
  G_{n \kappa} \left( r \right) 
  =
  \frac{ \frac{d}{dr} + \frac{\kappa}{r} }{\epsilon_{n \kappa} - V \left( r \right) + S \left( r \right) + 2mc^2}
  F_{n \kappa} \left( r \right).
\end{equation}
The normalization condition for the radial wave functions reads
\begin{equation}
  \label{eq12_}
  \int_0^{\infty} 
  \left[ F_{n \kappa}^2 \left( r \right) + G_{n \kappa}^2 \left( r \right) \right]
  dr 
  = 
  1.
\end{equation}
%
\subsection{The inverse Hamiltonian method}
\par
A Dirac Hamiltonian $ H_{\urm{D}} $ has the positive and negative eigenvalues.
The former and latter, respectively, correspond to the particle and anti-particle states.
Because of the existence of the negative-energy states, the ground-state energy is no longer the lowest eigenvalue of $ H_{\urm{D}} $.
Therefore, minimizing the energy naively leads to the variational collapse, i.e., the energy expectation value diverges into the negative infinite.
We introduce the so-called inverse Hamiltonian method~\cite{
  PhysRevC.82.057301}
to our DNN method in order to apply DNN for energy minimization to obtain the ground state.
\par
\begin{figure}[tb]
  \includegraphics[width=1.0\linewidth]{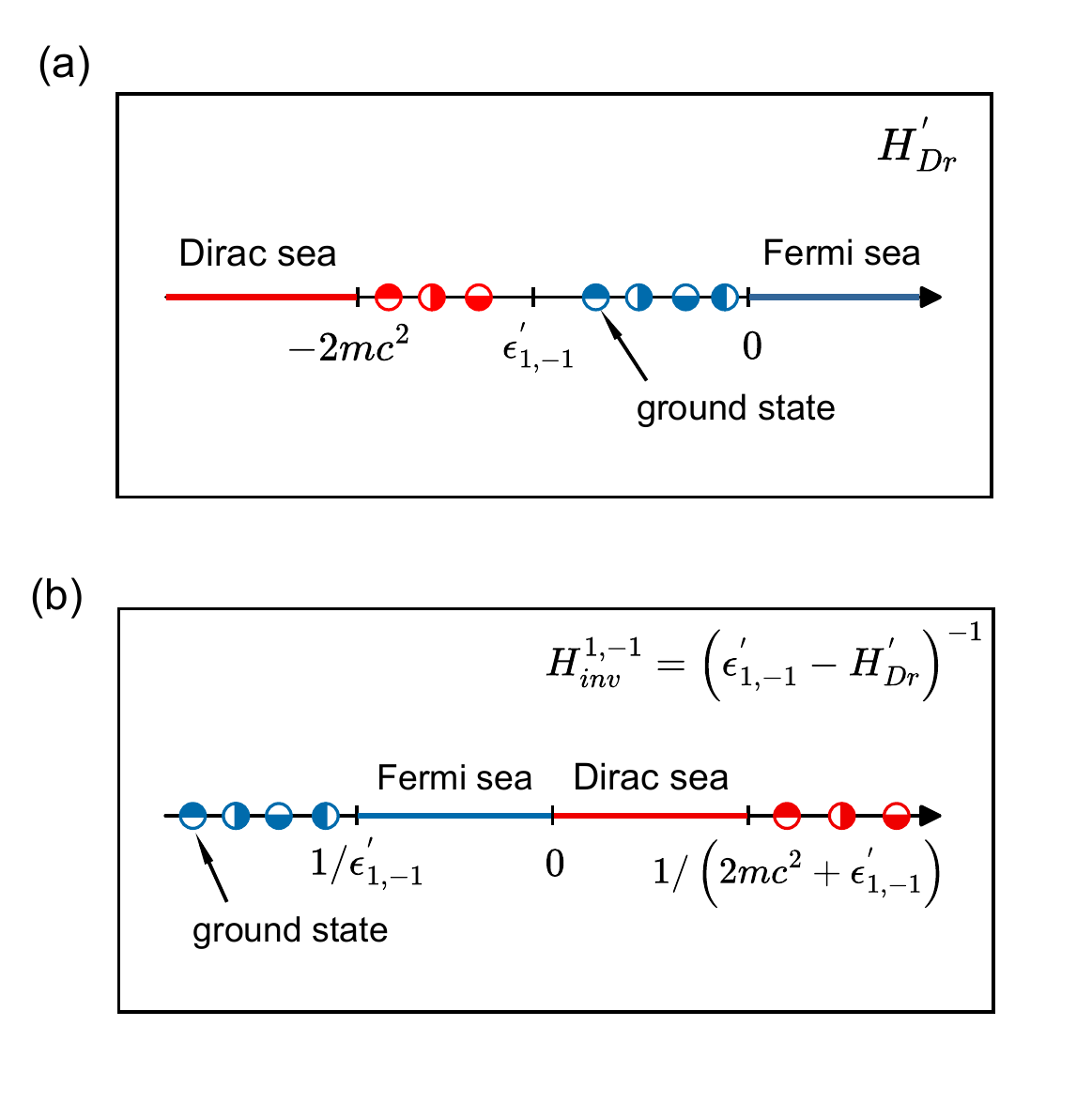}
  \caption{
    (a) Spectrum of eigenstates of the Dirac Hamiltonian $ H'_{\urm{D} r} $. 
    The half-filled blue circles denote the ground state and low-lying excited states, while the blue line denotes positive continuum states.
    The half-filled red circles and red line denote the bound states and continuum states in the Dirac sea, respectively.
    (b) Spectrum of eigenstates of the inverse Hamiltonian $ H_{\urm{inv}}^{1,-1} $~\cite{
      PhysRevC.82.057301}
    for the ground-state calculation. 
    The continuum states are reversed in the middle between bound states.
    The order among the bound states in both the Fermi sea and the Dirac sea does not change, respectively.
    The ground state is the lowest state in the spectrum. 
  }
  \label{i_H}
\end{figure}

The inverse Hamiltonian for given $ n $ and $ \kappa $ is defined by
\begin{equation}
  \label{eq12}
  H_{\urm{inv}}^{n \kappa} 
  =
  \left( \epsilon'_{n \kappa} - H'_{\urm{D} r} \right)^{-1}.
\end{equation}
The ground state is $ \kappa = -1 $ and $ n = 1 $;
thus, we choose an arbitrary value between the ground-state energy $ \epsilon_{1, -1} $
and the highest eigenvalue of negative energy states $ \left( \epsilon_{1, -1} - 2mc^2 \right) $ for
$ \epsilon'_{1, -1} $,
that is, $ \left( \epsilon_{1, -1} - 2mc^2 \right) < \epsilon'_{1, -1} < \epsilon_{1, -1} $
to obtain the ground-state energy.
The schematic figure of the spectrums of $ H'_{\urm{D} r} $ and $ H_{\urm{inv}}^{1, -1} $ is shown in Fig.~\ref{i_H}.
The lowest eigenstate of $ H_{\urm{inv}}^{1, -1} $ corresponds the ground state.
\par
Another benefit of applying the inverse Hamiltonian method is that both the ground and low-lying excited states can be calculated by varying $ \epsilon'_{n \kappa} $.
To calculate the state of quantum number $ n $ and $ \kappa $,
we set $ \epsilon'_{n \kappa} $ satisfying between
\begin{equation}
  \epsilon_{\left( n - 1\right)\kappa} 
  <
  \epsilon'_{ n \kappa}
  <
  \epsilon_{ n \kappa}.
\end{equation}
and calculate by Eq.~\eqref{eq12}
\begin{equation}
  \label{eq13}
  \epsilon_{n \kappa}
  =
  \epsilon'_{n \kappa}
  -
  \left(
    \min
    \frac{\brakket{\phi}{H_{\urm{inv}}^{n \kappa}}{\phi}}{\braket{\phi}{\phi}}
  \right)^{-1}.
\end{equation}
\par
Practically, $ \epsilon' $ is chosen slightly smaller than zero.
If the trial with another parameter $ \epsilon'' $ which is much smaller than $ \epsilon' $ returns the same eigenvalue,
we can safely assume the obtained state is the ground state.
If the test with $ \epsilon'' $ returns a lower eigenvalue,
we repeat the process until the ground state is obtained for two distinct trial parameters with a significant energy gap.
In the case with spherical symmetry, a more direct strategy is to count the number of nodes in the large component of wave functions, which is equal to $ n - 1 - l $.
Then, in a recursive approach, the state of quantum number $ n $ and $ \kappa $ can be obtained
by setting $ \epsilon'_{n \kappa} $ slightly greater than
$ \epsilon_{\left( n - 1 \right) \kappa} $.
The eigenvalue $\epsilon_{\left( n - 1 \right) \kappa} $ and $ \epsilon_{n \kappa} $ can be  splited as the highest and lowest eigenvalues in the spectrum of the inverse Hamiltonian.
Therefore, the inverse Hamiltonian method is a robust method even for highly dense spectra.
Note that the orthonormal method, which will be discussed in the next subsection, has proven in Ref.~\cite{
  PhysRevResearch.5.033189}
that it enables us to calculate even the degenerated spectrum, i.e., extremely dense spectra.
%
%
\subsection{The orthonormal method}
\par
Besides the inverse Hamiltonian method, we also extend the method to calculate low-lying excited states proposed in Ref.~\cite{
  PhysRevResearch.5.033189}
to obtain excited states of the Dirac equation.
The eigenfunctions satisfy the orthonormal condition
\begin{equation}
  \braket{\phi_{i \kappa}}{\phi_{n \kappa}}
  =
  \delta_{i, n}.
\end{equation}
For arbitrary states $ \ket{\phi} $ with a given $ \kappa $,
we can construct a state $ \ket{\phi'} $ that
is orthonormal to the eigenstates $ \ket{\phi_{i \kappa}} $ with
$ i = n_{\urm{min}} $, $ n_{\urm{min}} + 1 $, $ \ldots $, $ n $
\begin{equation}
  \label{eq31}
  \ket{\phi'}
  =
  \ket{\phi}
  -
  \sum_{i = n_{\uurm{min}}}^n
  \braket{\phi_{i \kappa}}{\phi}
  \ket{\phi_{i \kappa}}.
\end{equation}
Because the state $ \ket{\phi'} $ is orthonormal to eigenstates $ \ket{\phi_{i \kappa}} $ with lower principle quantum numbers from $ n_{\urm{min}} $ to $ n $,
the state of principle quantum number $ n + 1 $ is obtained by minimizing the energy variational
\begin{equation}
  \label{eq16}
  \epsilon_{\left(n + 1\right) \kappa}
  =
  \epsilon'_{1 \kappa}
  -
  \left(
    \min
    \frac{\brakket{\phi'}{H_{\urm{inv}}^{1 \kappa}}{\phi'}}{\braket{\phi'}{\phi'}}
  \right)^{-1}.
\end{equation}
Note that $ \ket{\phi'} $ is constructed in a recursive way for a given $ \kappa $,
so that wave functions of the lowest state cannot be obtained in this method.
Therefore, the lowest-energy state in Eq.~\eqref{eq31} is obtained by the inverse Hamiltonian method.
%
\section{Neural network models}
\label{Sect:III}
%
%
\subsection{Discretization of the Dirac equation}
\label{Sect:III,A}
\par
We define a trial wave function $ f_{n \kappa} $ by
\begin{equation}
  \label{eq17}
  f_{n \kappa}\left( r \right)
  =
  \frac{F_{n \kappa} \left( r \right)}{r}
\end{equation}
to give an accurate description of the asymptotic behavior near the origin.
The boundary condition reads $ f_{n \kappa} \left( 0 \right) = f_{n \kappa} \left( \infty \right) = 0 $.
\par
To perform numerical calculations, the radial coordinate
$ r \in \left[ 0, r_{\urm{max}} \right] $ is discretized with $ M + 1 $ meshes.
The Dirichlet boundary condition
$ f_{n \kappa} \left( 0 \right) = f_{n \kappa} \left( r_{\text{max}} \right) = 0 $
is imposed so that the actual number of mesh points used for DNN calculation is $ M - 1 $.
The discrete form of $ f_{n \kappa} \left( r \right) $ is a ($ M - 1 $)-dimensional vector
\begin{equation}
  f_{n \kappa} \left( r \right)
  \simeq
  \begin{pmatrix}
    \tilde{f}_{n \kappa 1} \\
    \tilde{f}_{n \kappa 2} \\
    \tilde{f}_{n \kappa 3} \\
    \vdots \\
    \tilde{f}_{n \kappa \left( M - 3 \right)} \\
    \tilde{f}_{n \kappa \left( M - 2 \right)} \\
    \tilde{f}_{n \kappa \left( M - 1 \right)}
  \end{pmatrix}.
\end{equation}
\par
Both the vector and scalar potentials are discretized as a ($ M - 1 $)-dimensional diagonal matrix:
\begin{widetext}
  \begin{equation}
    V \pm S
    \simeq
    \begin{pmatrix}
      \tilde{V}_1 \pm \tilde{S}_1 & 0 & 0 & \cdots & 0 & 0 & 0 \\
      0 & \tilde{V}_2 \pm \tilde{S}_2 & 0 & \cdots & 0 & 0 & 0 \\
      0 & 0 & \tilde{V}_3 \pm \tilde{S}_3 & \cdots & 0 & 0 & 0 \\
      \vdots & \vdots & \vdots & \ddots & \vdots & \vdots & \vdots\\
      0 & 0 & 0 & \cdots & \tilde{V}_{M - 3} \pm \tilde{S}_{M - 3} & 0 & 0 \\
      0 & 0 & 0 & \cdots & 0 & \tilde{V}_{M - 2} \pm \tilde{S}_{M - 2} & 0 \\
      0 & 0 & 0 & \cdots & 0 & 0 & \tilde{V}_{M - 1} \pm \tilde{S}_{M - 1} 
    \end{pmatrix}
  \end{equation}
\end{widetext}
with $ \tilde{V}_i = V \left( r_i \right) $ and $ \tilde{S}_i = S \left( r_i \right) $.
The form of the differential operator depends on the distribution of mesh points, which will be discussed in Sec.~\ref{Sect:IV,B} and Sec.~\ref{Sect:V,B}.
Then, the Dirac Hamiltonian $ H'_{\urm{D} r} $ is expressed by a ($ 2M - 2 $)-dimensional matrix,
which is used to construct the inverse Hamiltonian $ H_{\urm{inv}}^{n \kappa} $ numerically.
%
%
\subsection{Design of the neural network}
\label{Sect:III,B}
\par
\begin{figure}[tb] 
  \centering
  \includegraphics[width=1.0\linewidth]{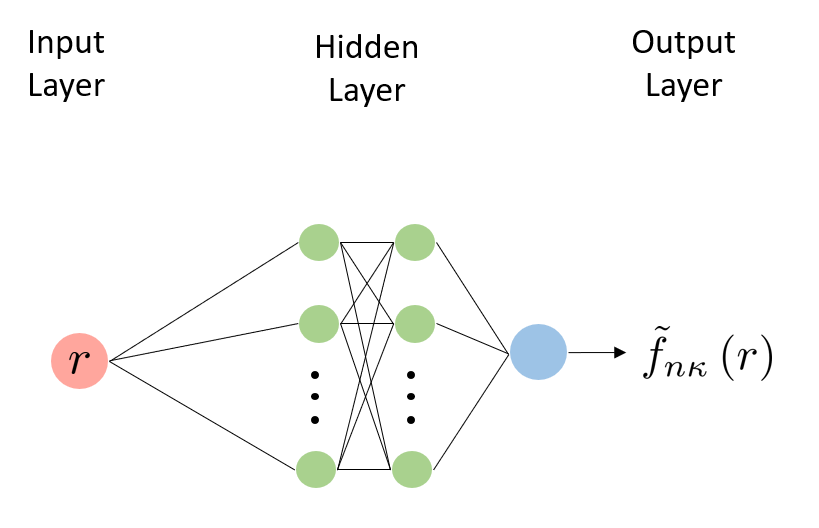}
  \caption{
    Schematic figure of the DNN structure.
    The input layer contains one unit (red), where the input is the radial coordinate $ r $.
    The output layer contains one unit (blue), where the output is $ \tilde{f}_{n \kappa} \left( r \right) $.
    The hidden layers are composed of two fully connected layers, each of which contains 16 units (green).}
  \label{DNN}
\end{figure}

We construct a fully connected neural network for the trial wave function $ f_{n \kappa} \left( r \right) $
on the \textsc{TensorFlow}~\cite{Tensorflow}.
Figure~\ref{DNN} shows the structure of the DNN, which contains one input unit and one output unit.
It is confirmed that two hidden layers, each of which contains 16 units, are sufficient to generate faithful results, which is similar to our previous work~\cite{wang2024neural}.
The ``softplus'' function 
\begin{equation}
  \softplus
  \left( x \right)
  =
  \log
  \left(
    1
    +
    e^x
  \right)
\end{equation}
is used as the activation function.
The parameters of the DNN are updated by the \textsc{Adam} optimizer~\cite{
  DBLP:journals/corr/KingmaB14} with a fixed learning rate of 0.001.
Regularization techniques are not applied, as no overfitting is observed during the trainings.
\par
Throughout the training process, all radial coordinates are treated as one batch sending into the DNN and the mini-batch technique is not used.
The wave function $ F_{n \kappa} $ is calculated by the DNN output $ f_{n \kappa} $ by Eq.~\eqref{eq17},
and then it is used to calculate $ G_{n \kappa} $ by Eq.~\eqref{eq9}.
For the ground state, the loss function $ \epsilon_{n \kappa} $ is calculated by Eq.~\eqref{eq13},
while either Eq.~\eqref{eq13} or \eqref{eq16} is used for excited states.
According to Eq.~\eqref{eq9}, $ \epsilon_{n \kappa} $ is needed to calculate $ G_{n \kappa} $.
For the $ n $-th epoch calculation, the $ \epsilon_{n \kappa} $ to calculate $ G_{n \kappa} $ is from the ($ n - 1 $)-th epoch,
while the initial $ \epsilon_{n \kappa} $ is set equal to $ \epsilon_{n \kappa}' $ for the first epoch.
%
%
\section{Results of the Hydrogen Atom}
\label{Sect:IV}
%
%
\subsection{The exact results}
\label{Sect:IV,A}
\par
We apply our DNN method for the hydrogen atom.
The vector and scalar potentials, respectively, read
\begin{subequations}
  \begin{align}
    \label{eq38}
    V \left( r \right)  
    & =
      -
      \frac{1}{r},\\
    S \left( r \right)
    & =
      0,
  \end{align}
\end{subequations}
where the Hartree atomic unit is used, i.e., $ \hbar = 4 \pi \epsilon_0 = e^2 = m_e = 1 $.
\par
The wave functions and energy are analytically known as~\cite{
  sakurai1967advanced}
\begin{subequations}
  \begin{align}
    F_{n \kappa}
    & =
      e^{-\rho} 
      \rho^s 
      \sum_{m = 0}^{n - 1/2 - j} 
      a_m 
      \rho^m, \\
    G_{n \kappa}
    & = 
      e^{-\rho} 
      \rho^s 
      \sum_{m = 0}^{n - 1/2 - j} 
      b_m 
      \rho^m, \\
    E_{n \kappa}
    & = 
      \frac{c^2}{\sqrt{1 + \frac{Z^2 \alpha^2}{\left( n - j - \frac{1}{2} + \sqrt{\left( j + \frac{1}{2} \right)^2 - Z^2 \alpha^2}\right)^2}}},
  \end{align}
\end{subequations}
with
\begin{subequations}
  \begin{align}
    s 
    & =
      \sqrt{\kappa^2 - 1/c^2},\\
    \rho 
    & = 
      \frac{\sqrt{c^4 - E_{n \kappa}^2}}{c}
      r,\\
    \mu 
    & = 
      \sqrt{\frac{c^2 - E_{n \kappa}^2}{c^2 + E_{n \kappa}^2}},
  \end{align}
\end{subequations}
and the speed of light $ c $.
The normalization coefficients $ a_q $ and $ b_q $ are calculated by
\begin{equation}
  C_q
  =
  \frac{a_q}{\frac{s + q -\kappa}{\mu} + \frac{1}{c}}
  =
  \frac{b_q}{s + q + \kappa - \frac{1}{c \mu}},
\end{equation}
where $ C_q $ satisfies the recurrence equation
\begin{equation}
  C_q 
  = 
  \frac{2 \left( q - n + j - \frac{1}{2} \right)}{q \left( q + 2s \right)}
  C_{q-1}.
\end{equation}
%
%
\subsection{Discrete representations with the log mesh}
\label{Sect:IV,B}
\par
The potential changes rapidly in the central region (small $ r $).
In order to calculate the central region precisely,
we define $ x $ by
\begin{equation}
  \label{eq29}
  r = e^x - e^{x_0},
\end{equation}
and $ x $ is discretized uniformly with the size $ \Delta x $ with $ x_0 $ as the small boundary point.
Note that the term $ e^{x_0} $ is introduced to impose the boundary condition
$ f_{n \kappa} \left( {r = 0} \right) = 0 $ properly, 
as the following Eq.~\eqref{eq30} requiring
$ f_{n \kappa} \left( r \left( x_0 \right) \right) = 0 $.
The first-order differential operator for the log mesh reads
\begin{equation}
  \frac{\partial}{\partial r} 
  =
  \frac{1}{e^x}
  \frac{\partial}{\partial x},
\end{equation}
where the corresponding derivative matrix reads
\begin{widetext}
\begin{equation}
  \label{eq30}
  \frac{\partial}{\partial r} 
  \simeq
  \frac{1}{2 \Delta x}  
  \begin{pmatrix}
    0 & \frac{1}{e^{x_1}} & 0 & \cdots & 0 & 0 & 0 \\
    - \frac{1}{e^{x_2}} & 0 & \frac{1}{e^{x_2}} & \cdots & 0 & 0 & 0 \\
    0 & - \frac{1}{e^{x_3}} & 0 & \cdots & 0 & 0 & 0 \\
    \vdots & \vdots & \vdots & \ddots & \vdots & \vdots & \vdots \\
    0 & 0 & 0 & \cdots & 0 & \frac{1}{e^{x_{\left( M - 3 \right)}}} & 0 \\
    0 & 0 & 0 & \cdots & - \frac{1}{e^{x_{\left( M - 2 \right)}}} & 0 & \frac{1}{e^{x_{\left( M - 2 \right)}}} \\
    0 & 0 & 0 & \cdots & 0 & - \frac{1}{e^{x_{\left( M - 1 \right)}}} & 0
  \end{pmatrix}.
\end{equation}
\end{widetext}
%
%
\subsection{DNN results}
\label{Sect:IV,C}
\par
In this section, we use $ M = 1700 $ as the number of mesh points.
For the orthonormal method, all states are calculated within a
$ e^{4.9} - e^{-10} \approx 100 \, \mathrm{a.u.} $
box to impose the orthonormal condition,
while the box sizes of the inverse Hamiltonian are $ 20 $, $ 40 $, $ 40 $, $ 60 $, $ 90 $, $ 100 \, \mathrm{a.u.} $ for states from $ n = 1 $ to $ 6 $, respectively.
\par
The energies of the ground state and first five excited states for $ \kappa = -1 $ are listed in Table~\ref{Hkappa-1}.
The inverse Hamiltonian method reaches at least $ 1 \times 10^{-4} $ accuracy and the error does not depend on $ n $ except for $ n = 1 $.
The relative error of the orthonormal method increases with the increase of $ n $.
This is because as $ n $ increases,
more lower-energy states are used to construct $ \ket{\phi'} $ from Eq.~\eqref{eq31},
so that the error of lower-energy state data accumulates.
\par
\begin{figure*}[tb]
  \begin{minipage}{0.32\textwidth}
    \centering
    \includegraphics[width=1.0\linewidth]{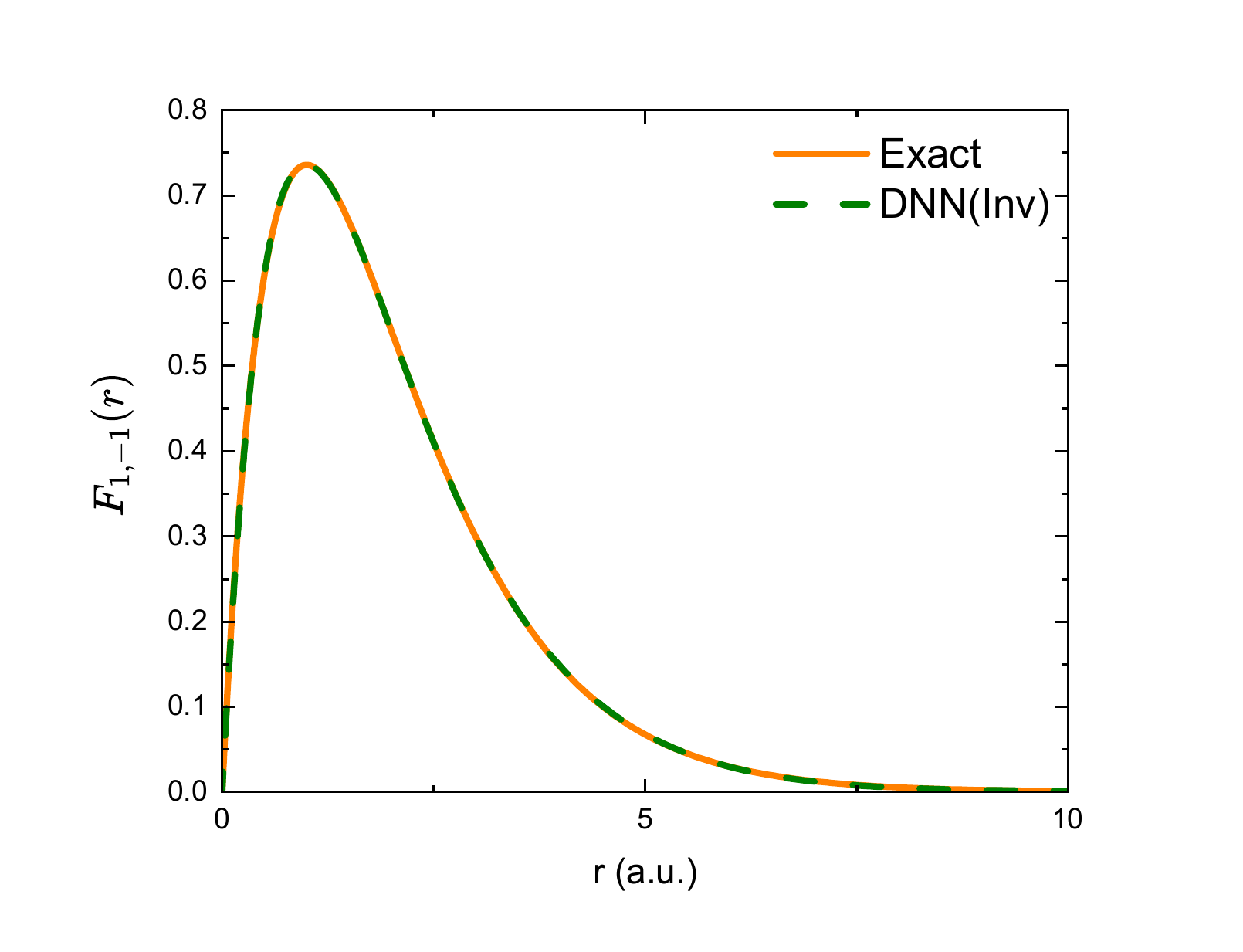}
  \end{minipage}
  \hfill
  \begin{minipage}{0.32\textwidth}
    \centering
    \includegraphics[width=1.0\linewidth]{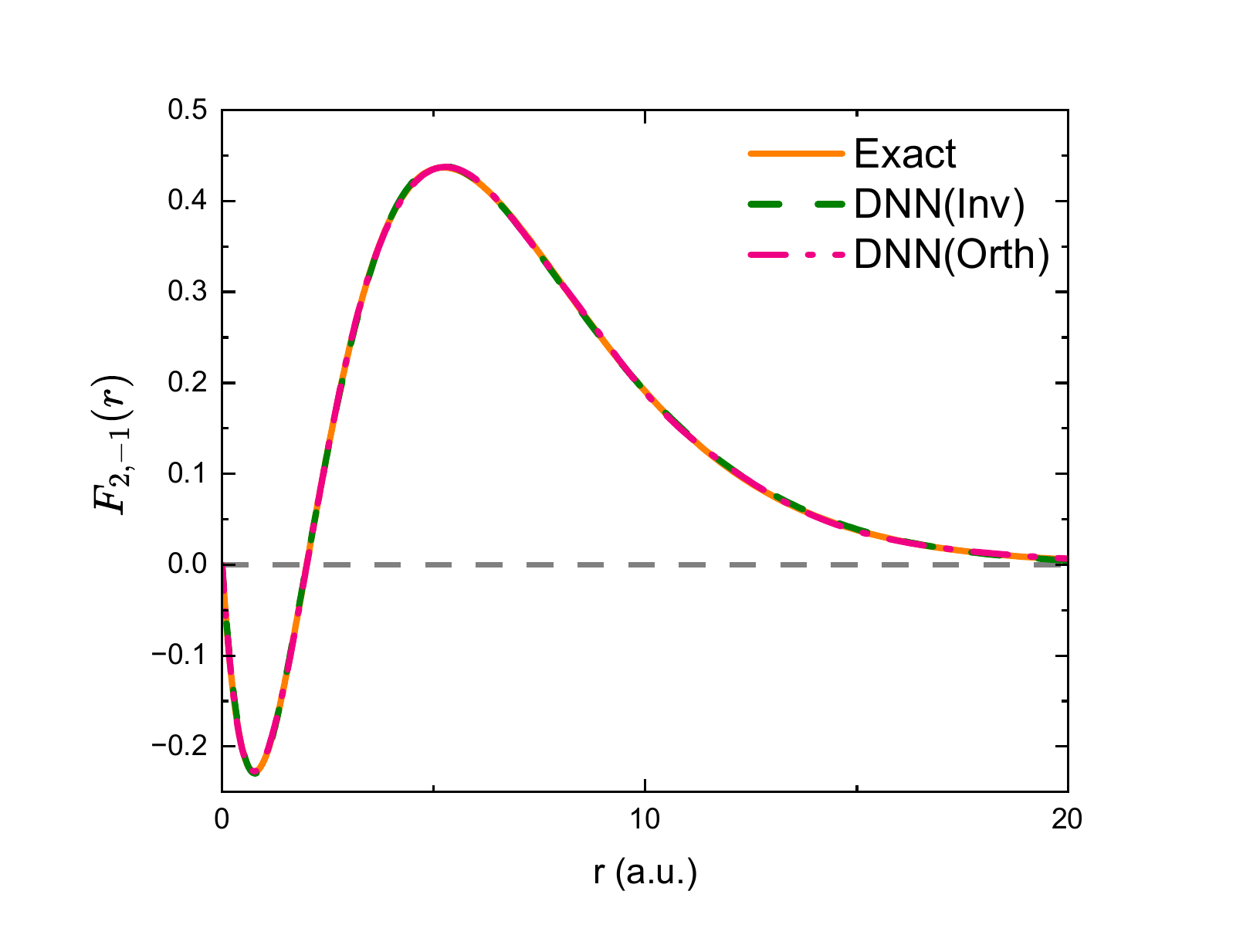}
  \end{minipage}
  \hfill
  \begin{minipage}{0.32\textwidth}
    \centering
    \includegraphics[width=1.0\linewidth]{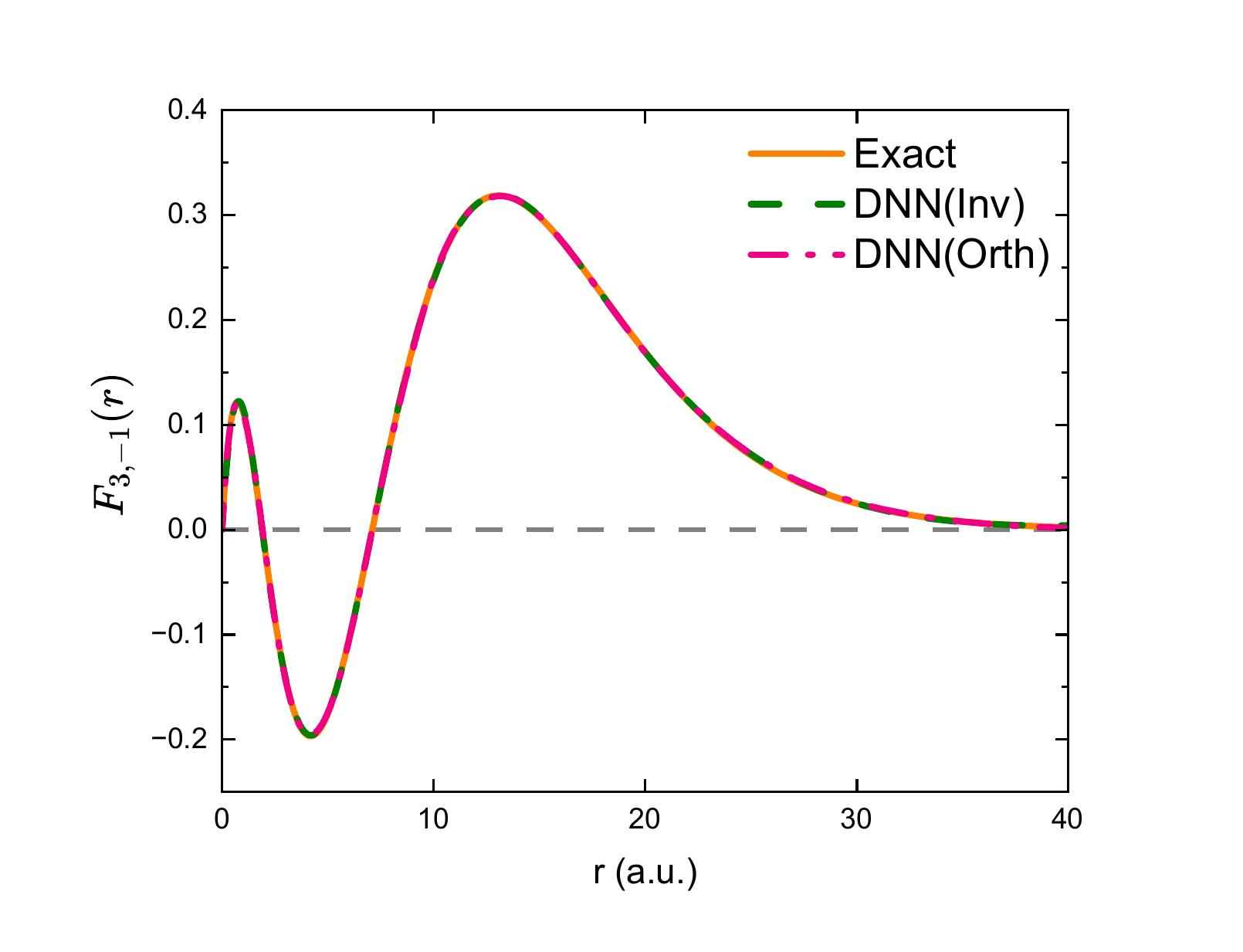}
  \end{minipage} \\
  \begin{minipage}{0.32\textwidth}
    \centering
    \includegraphics[width=1.0\linewidth]{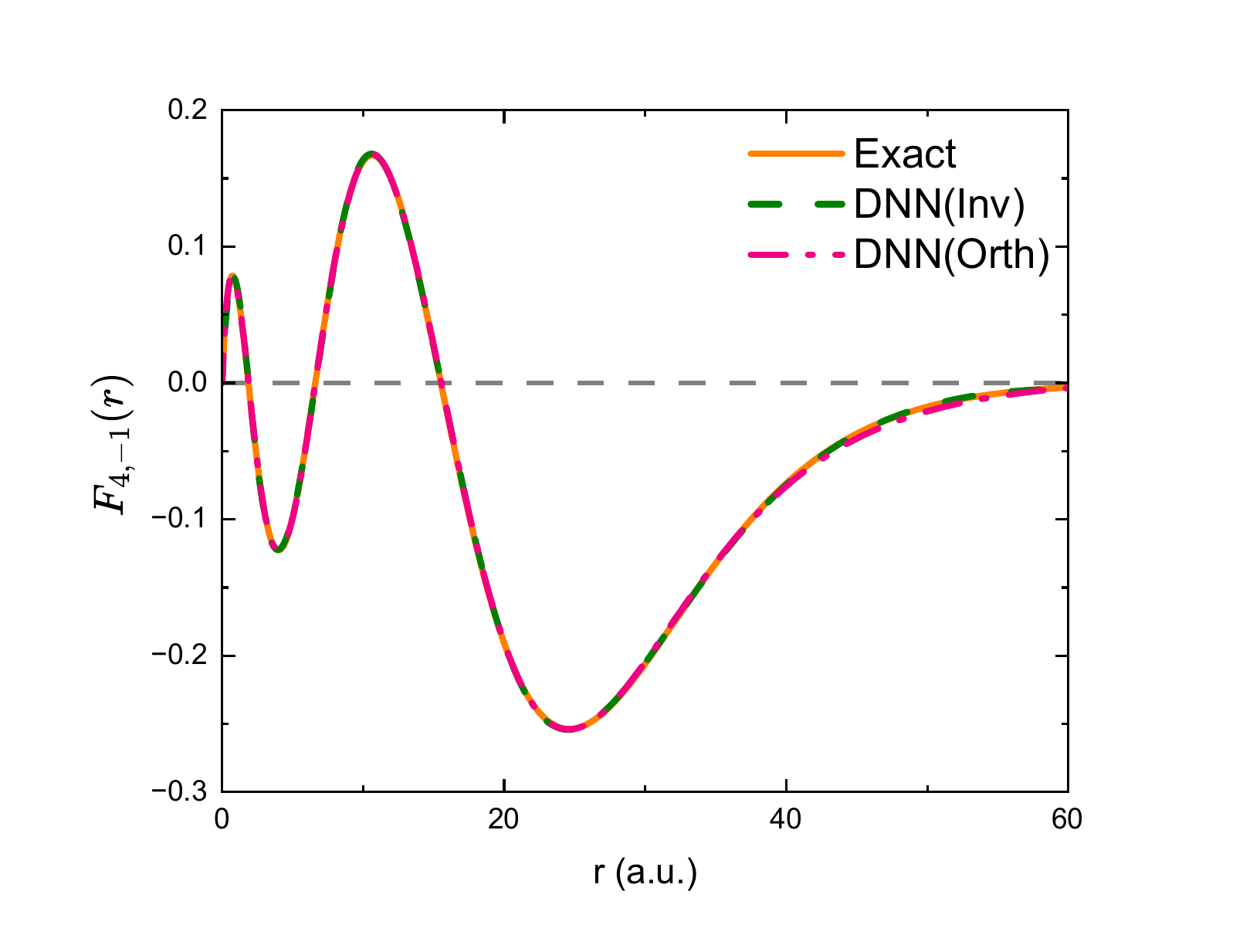}
  \end{minipage}
  \hfill
  \begin{minipage}{0.32\textwidth}
    \centering
    \includegraphics[width=1.0\linewidth]{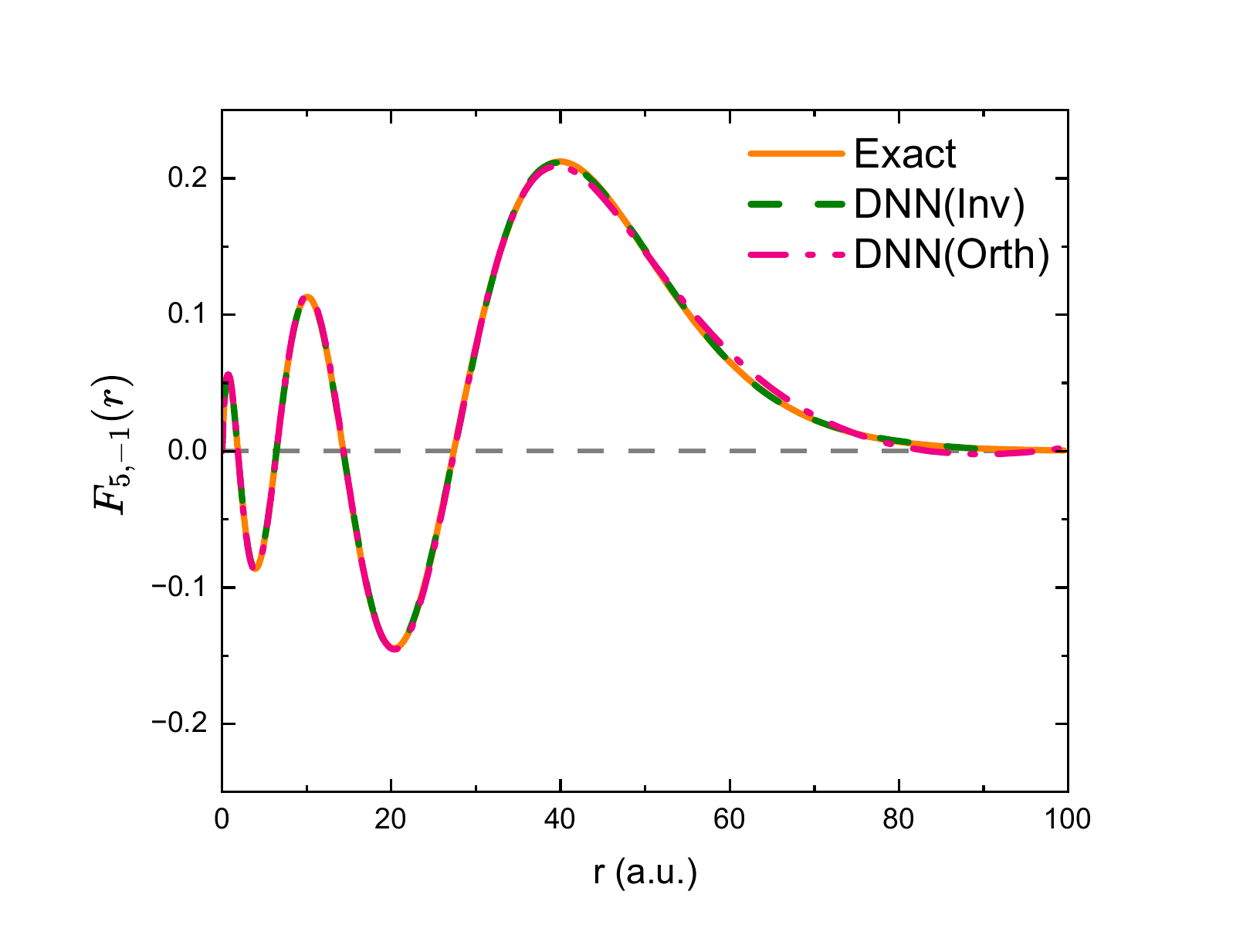}
  \end{minipage}
  \hfill
  \begin{minipage}{0.32\textwidth}
    \centering
    \includegraphics[width=1.0\linewidth]{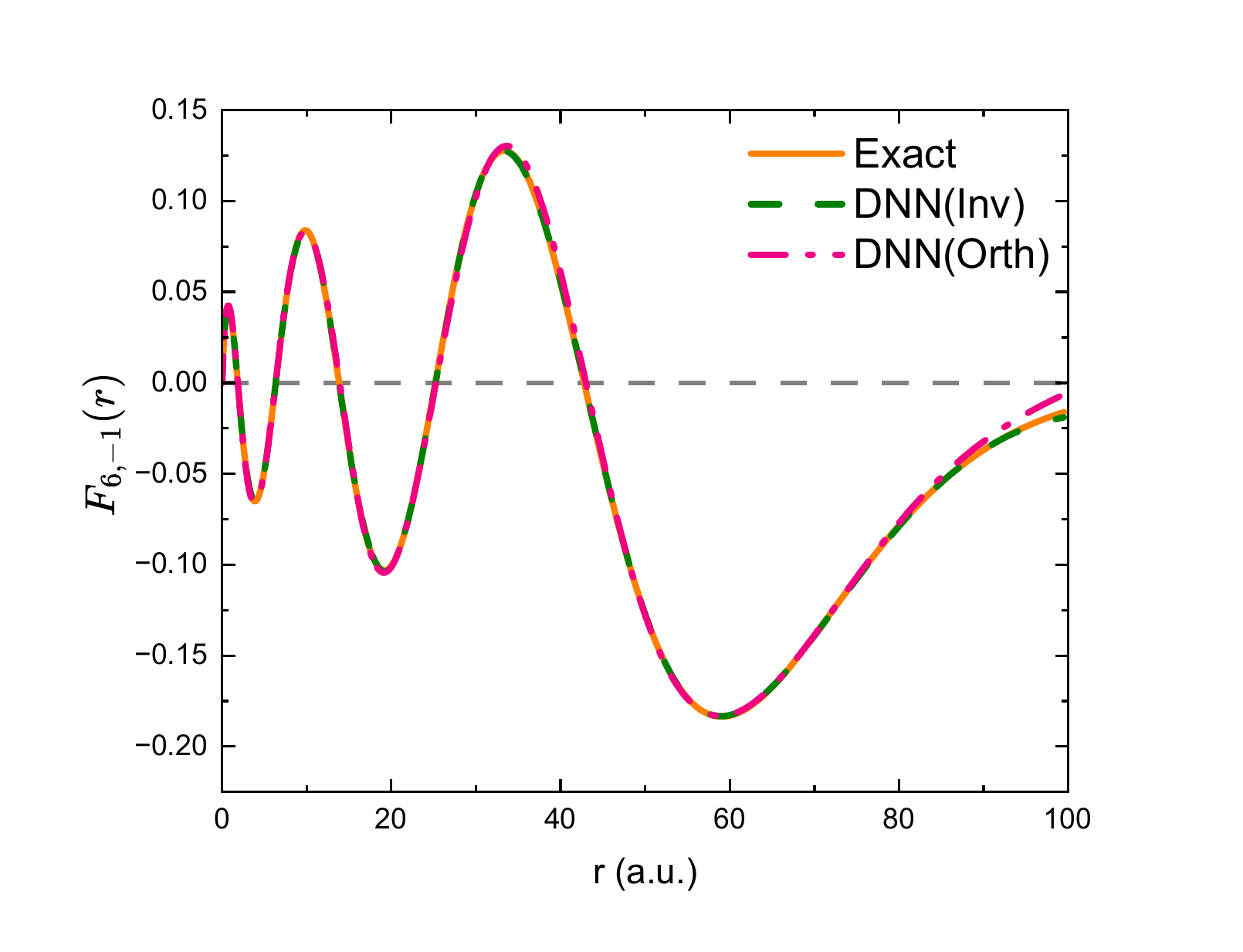}
  \end{minipage}
  \caption{
    The $ F $-component of Dirac wave functions with $ \kappa = -1$ of the hydrogen atom from $ n = 1 $ to $ 6 $.
    The DNN results from the inverse Hamiltonian method and the orthonormal one are shown as ``DNN(Inv)'' and ``DNN(Orth)'' in dash and dash-dotted lines, respectively.
    For comparison, the analytical results are also shown in the solid lines.
    Both the inverse Hamiltonian method and the orthonormal one reproduce results accurately.}
  \label{F-1}
\end{figure*}
\begin{figure*}[tb]
  \begin{minipage}{0.32\textwidth}
    \centering
    \includegraphics[width=1.0\linewidth]{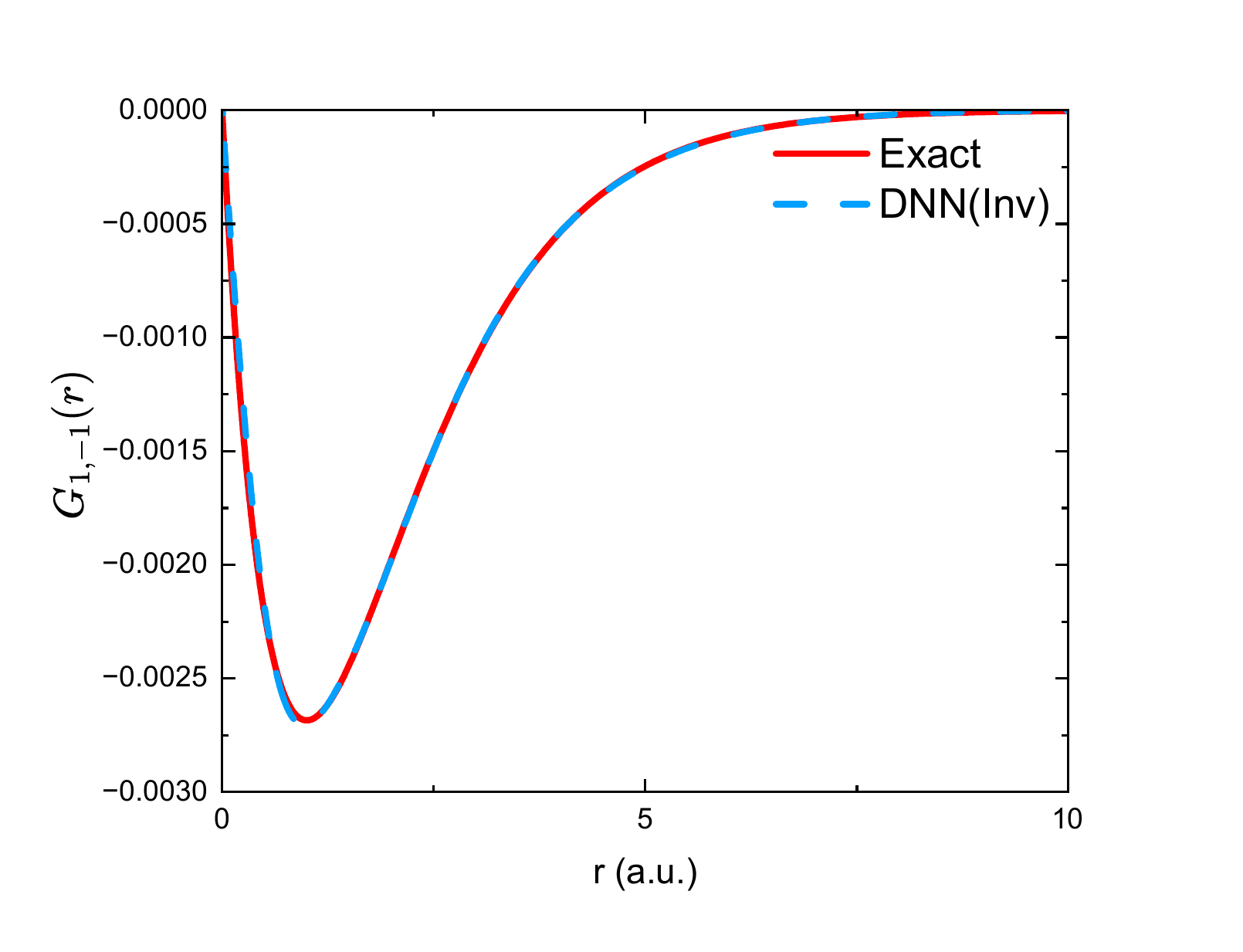}
  \end{minipage}
  \hfill
  \begin{minipage}{0.32\textwidth}
    \centering
    \includegraphics[width=1.0\linewidth]{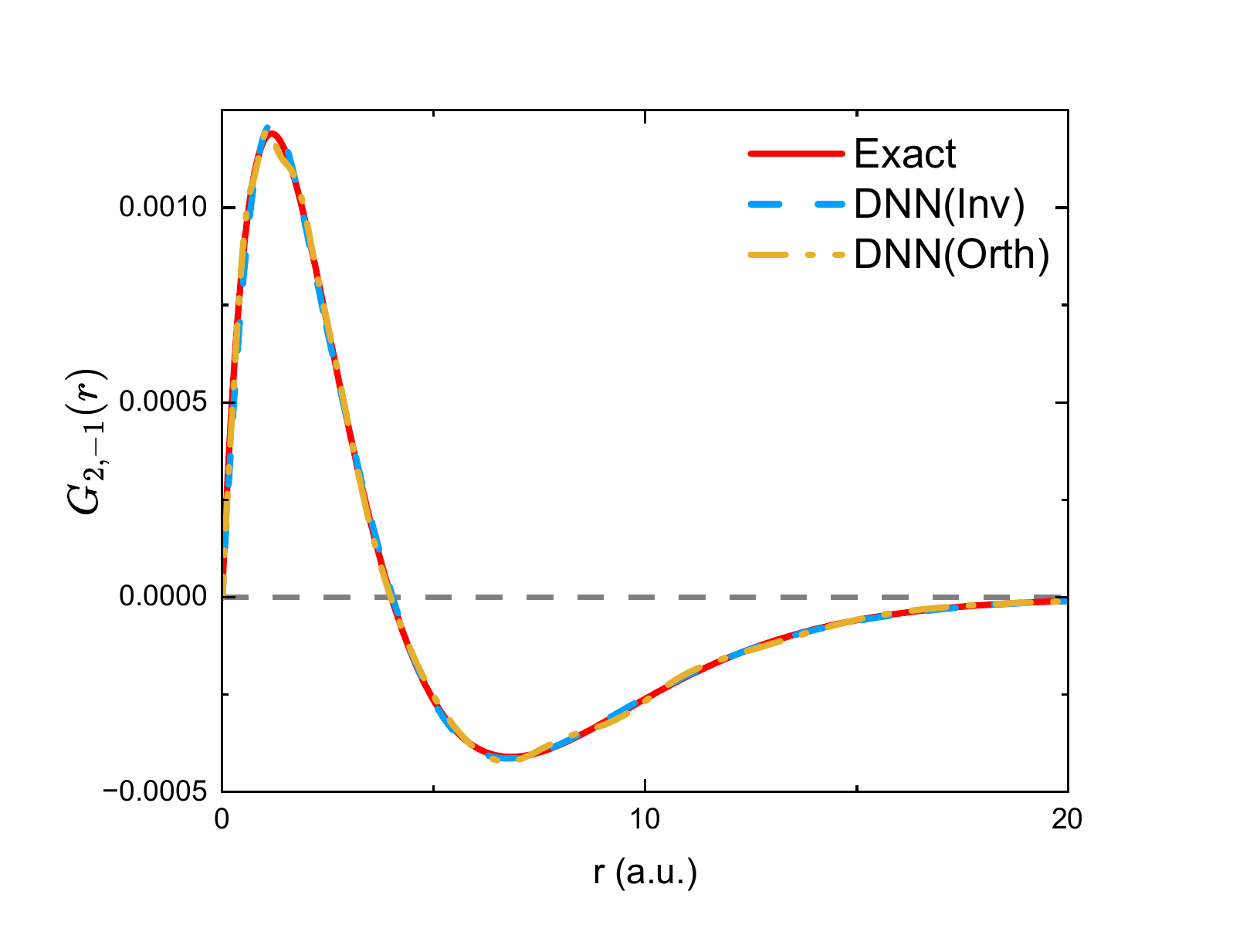}
  \end{minipage}
  \hfill
  \begin{minipage}{0.32\textwidth}
    \centering
    \includegraphics[width=1.0\linewidth]{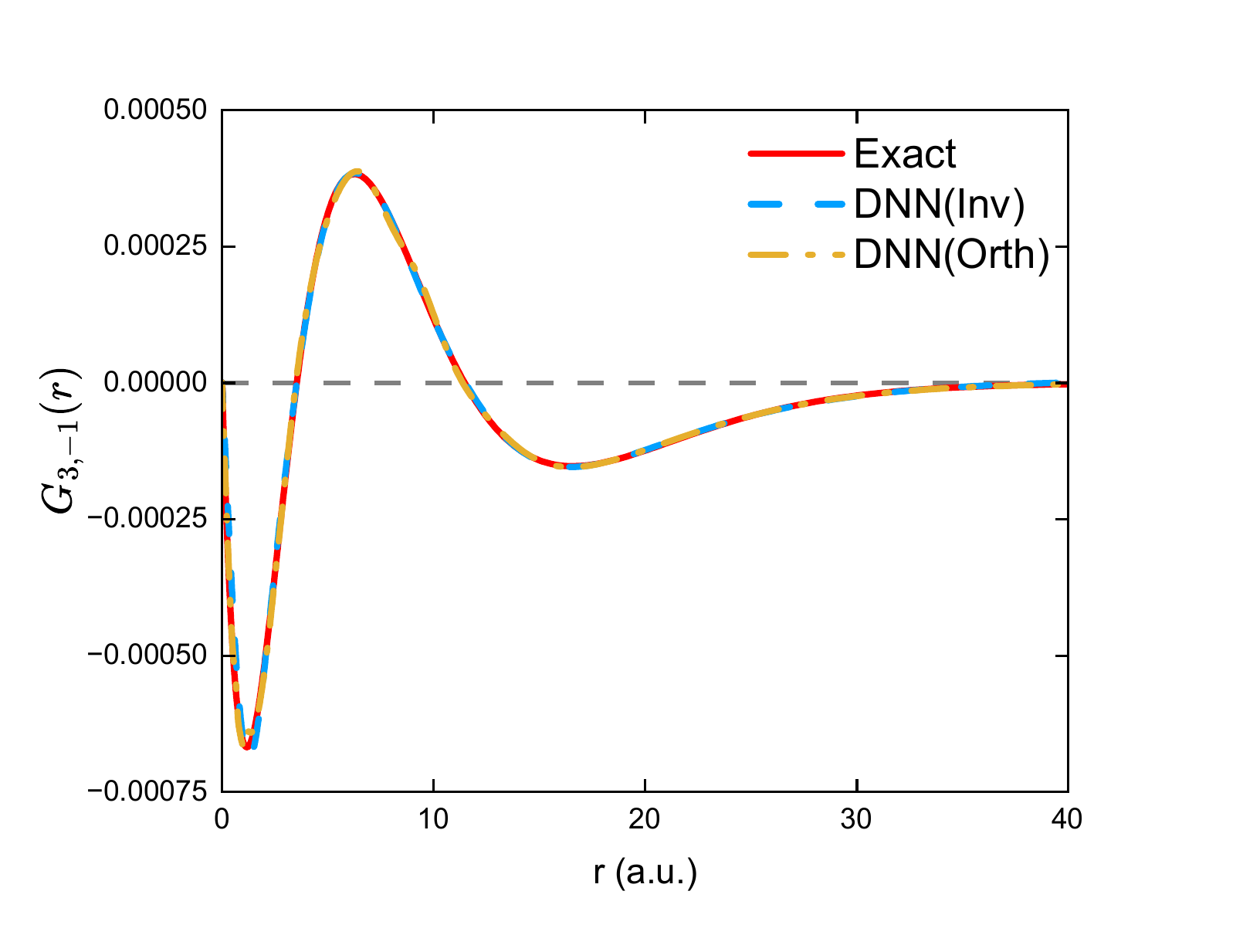}
  \end{minipage} \\
  \begin{minipage}{0.32\textwidth}
    \centering
    \includegraphics[width=1.0\linewidth]{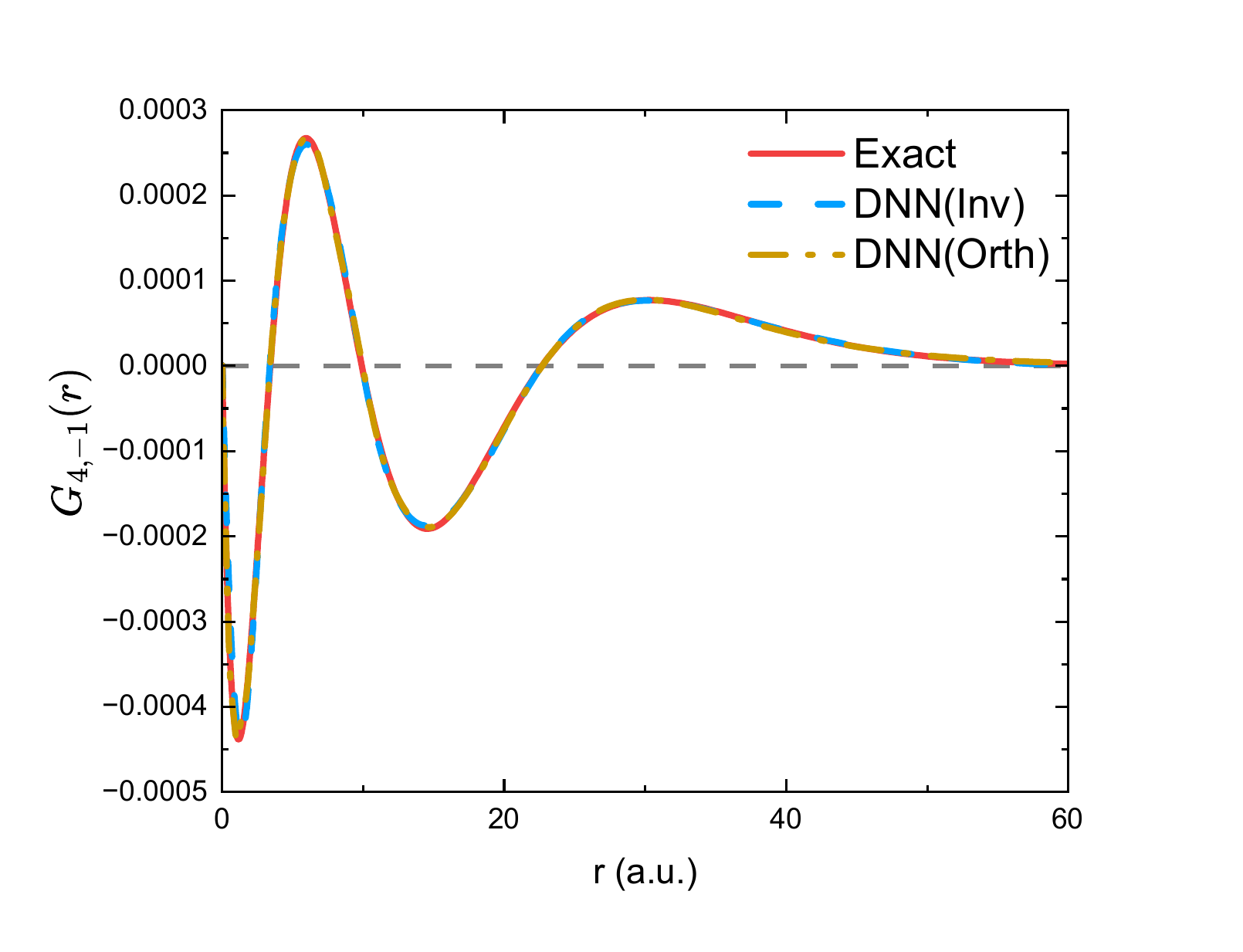}
  \end{minipage}
  \hfill
  \begin{minipage}{0.32\textwidth}
    \centering
    \includegraphics[width=1.0\linewidth]{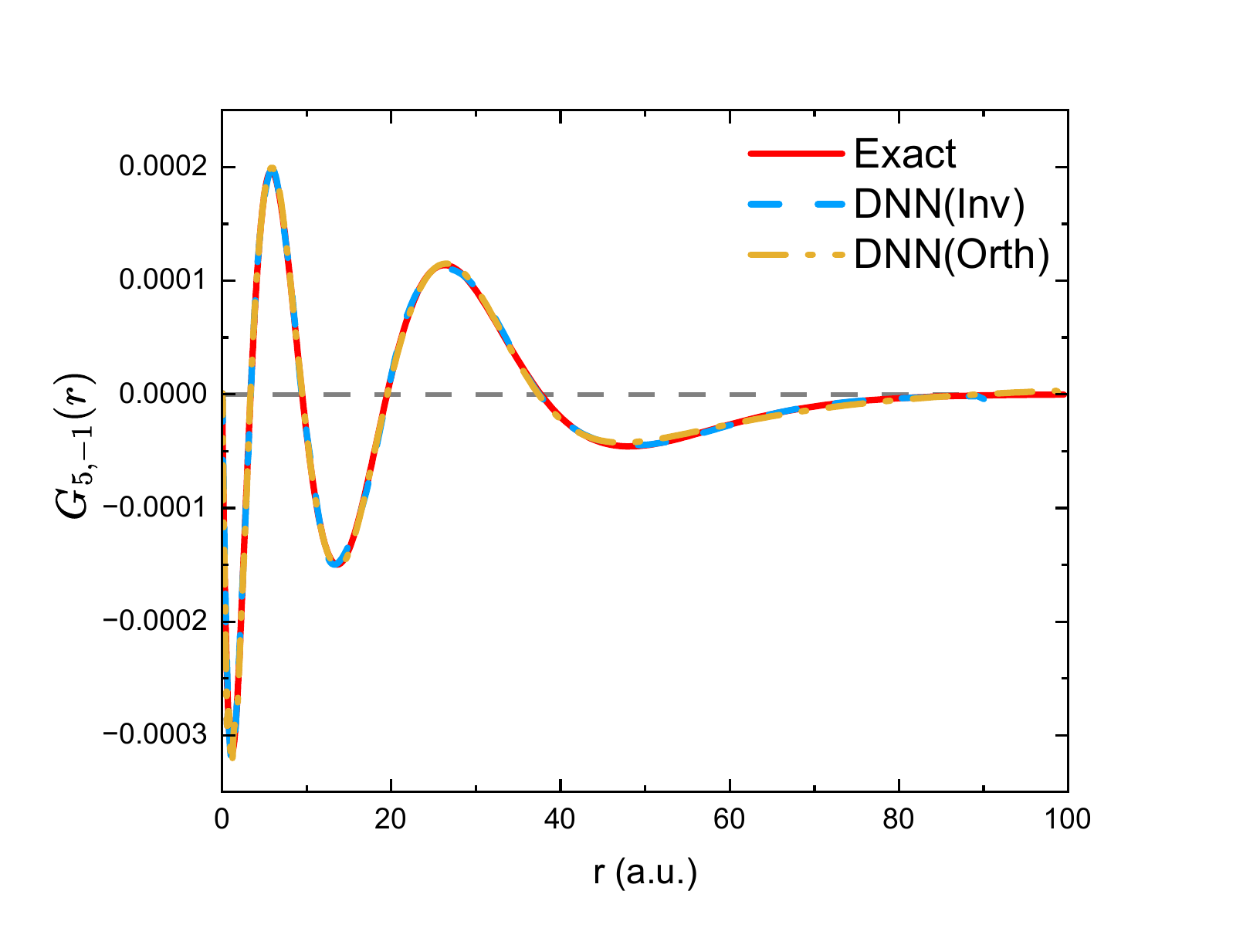}
  \end{minipage}
  \hfill
  \begin{minipage}{0.32\textwidth}
    \centering
    \includegraphics[width=1.0\linewidth]{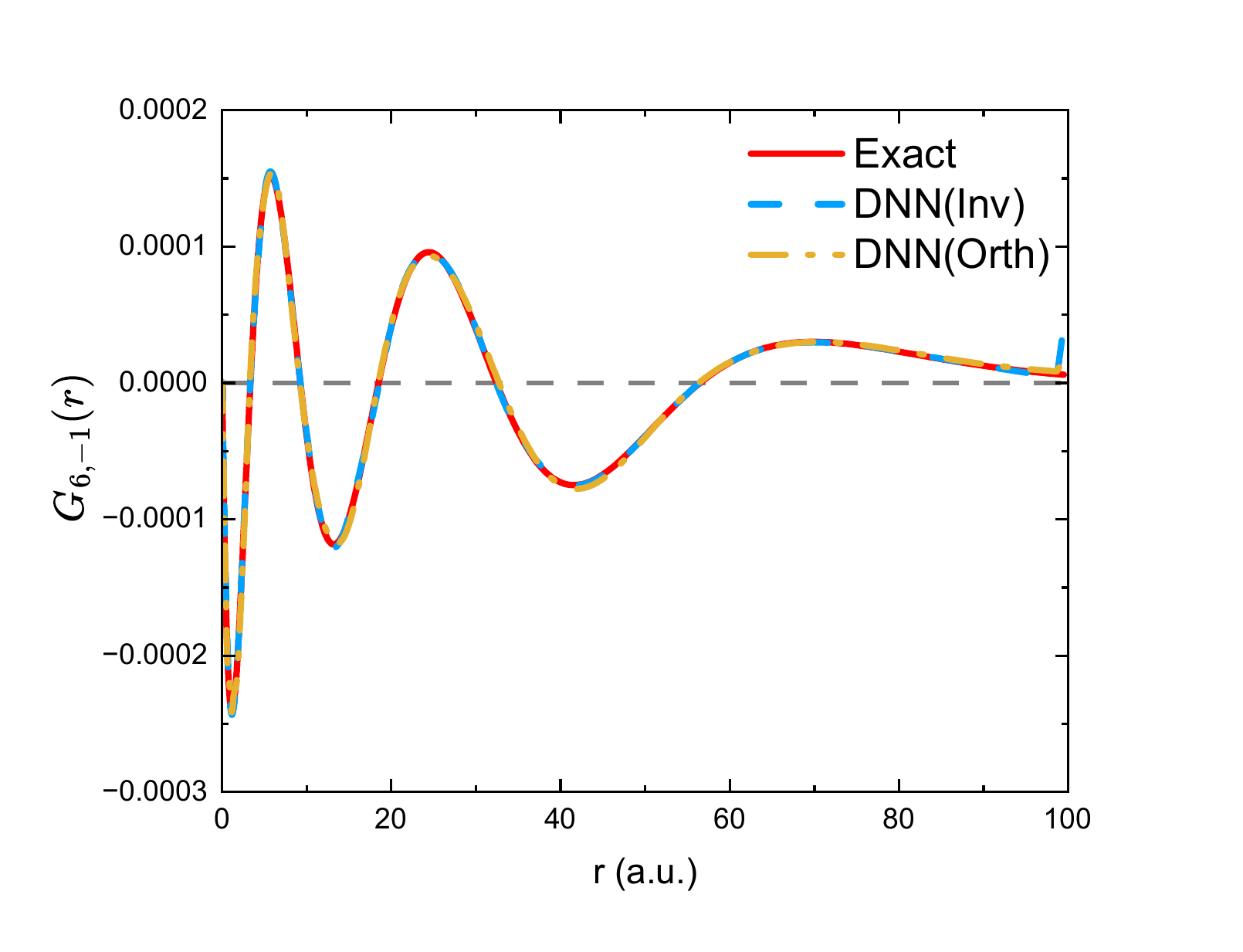}
  \end{minipage}
  \caption{
    Same as Fig.~\ref{F-1}, but for the $ G $-component.}
  \label{G-1}
\end{figure*}
\begin{figure*}[tb]
  \begin{minipage}{0.32\textwidth}
    \centering
    \includegraphics[width=1.0\linewidth]{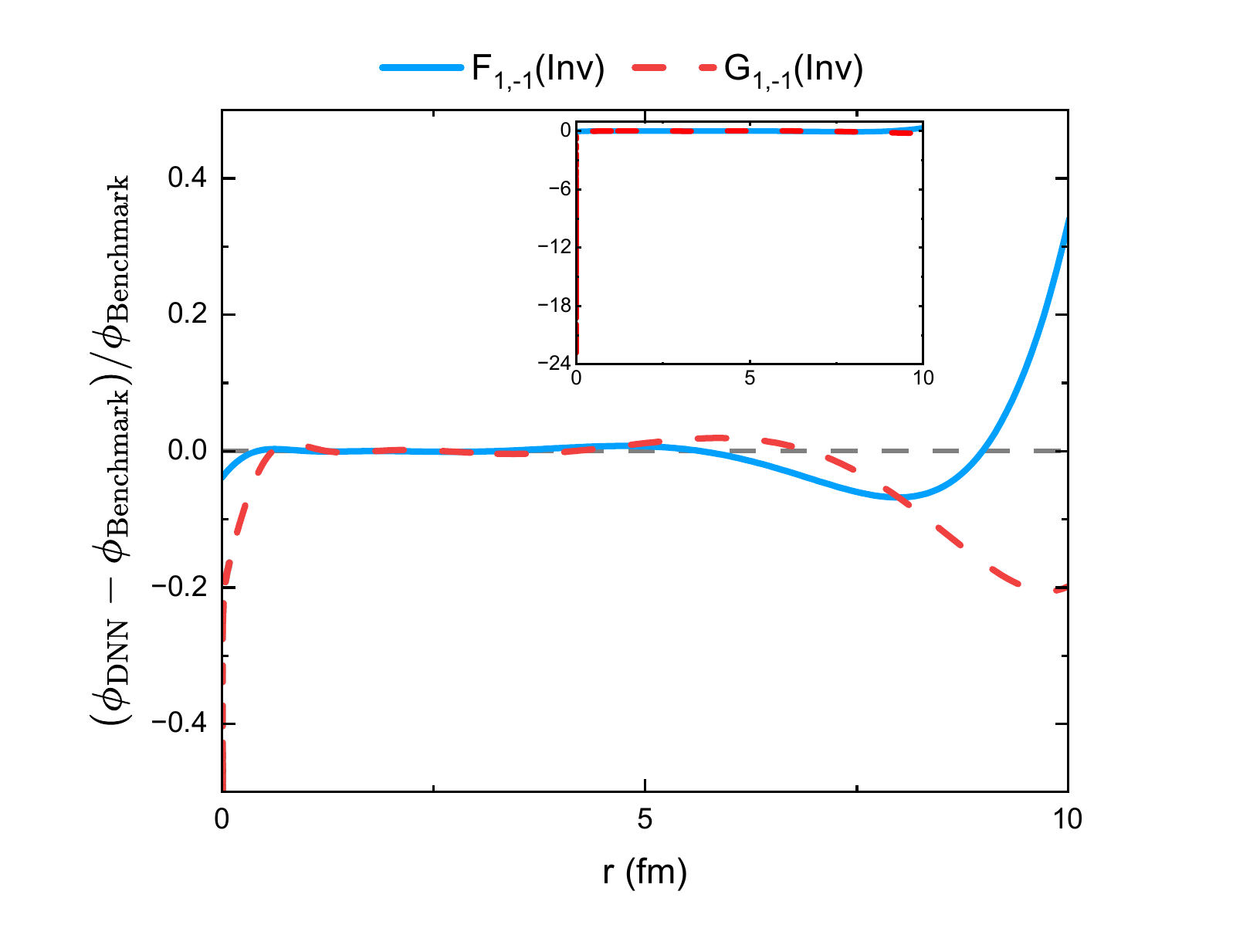}
  \end{minipage}
  \hfill
  \begin{minipage}{0.32\textwidth}
    \centering
    \includegraphics[width=1.0\linewidth]{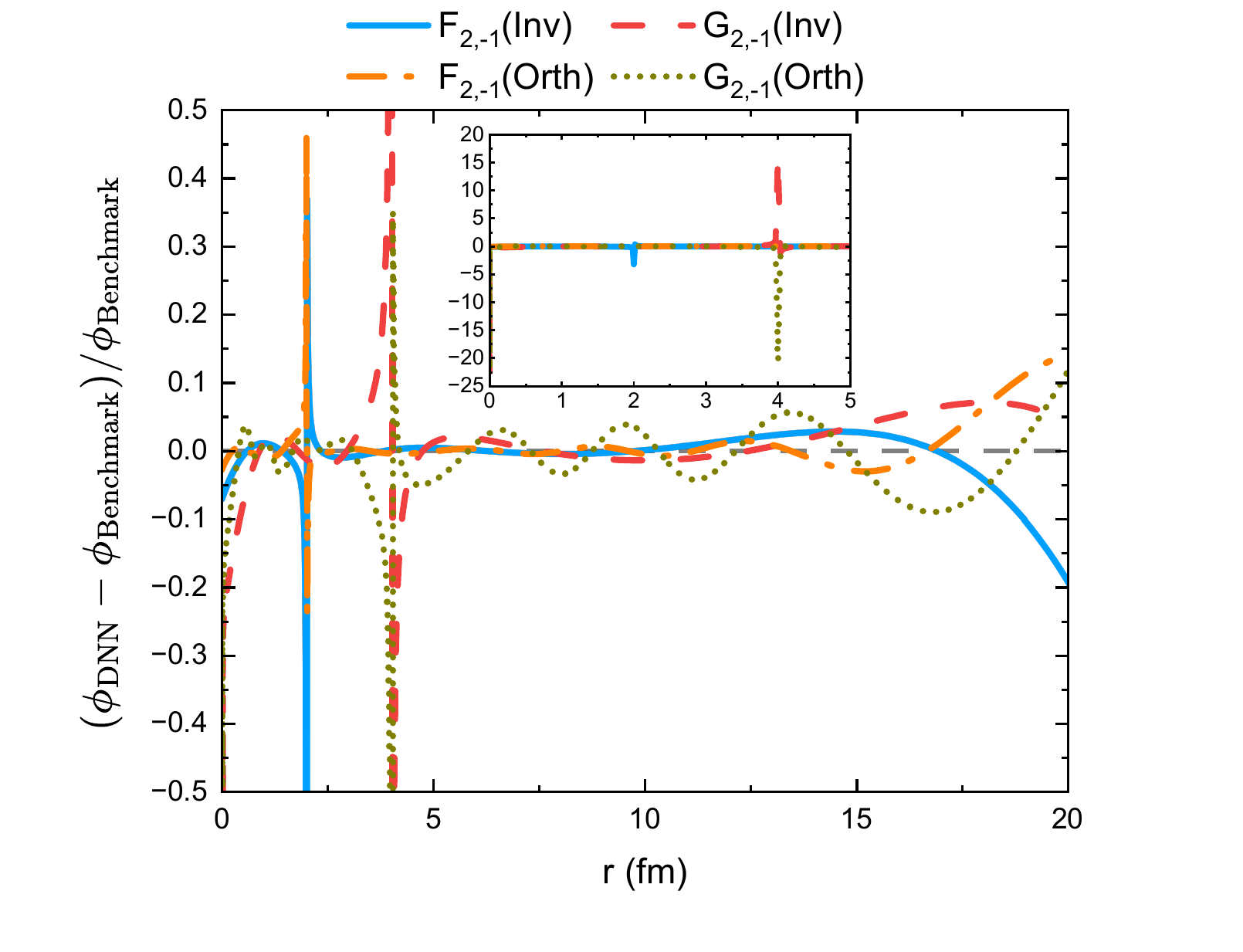}
  \end{minipage}
  \hfill
  \begin{minipage}{0.32\textwidth}
    \centering
    \includegraphics[width=1.0\linewidth]{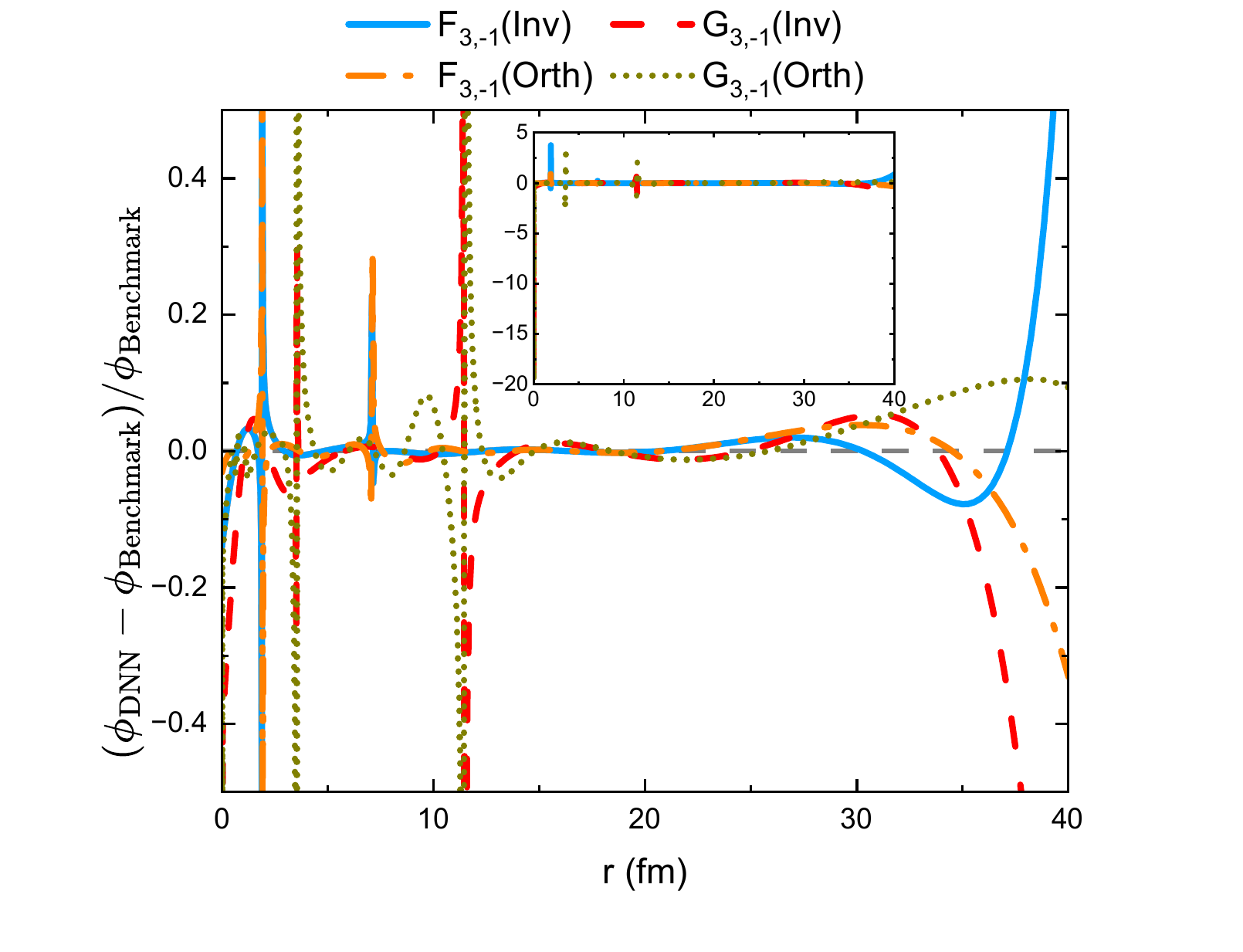}
  \end{minipage} \\
  \begin{minipage}{0.32\textwidth}
    \centering
    \includegraphics[width=1.0\linewidth]{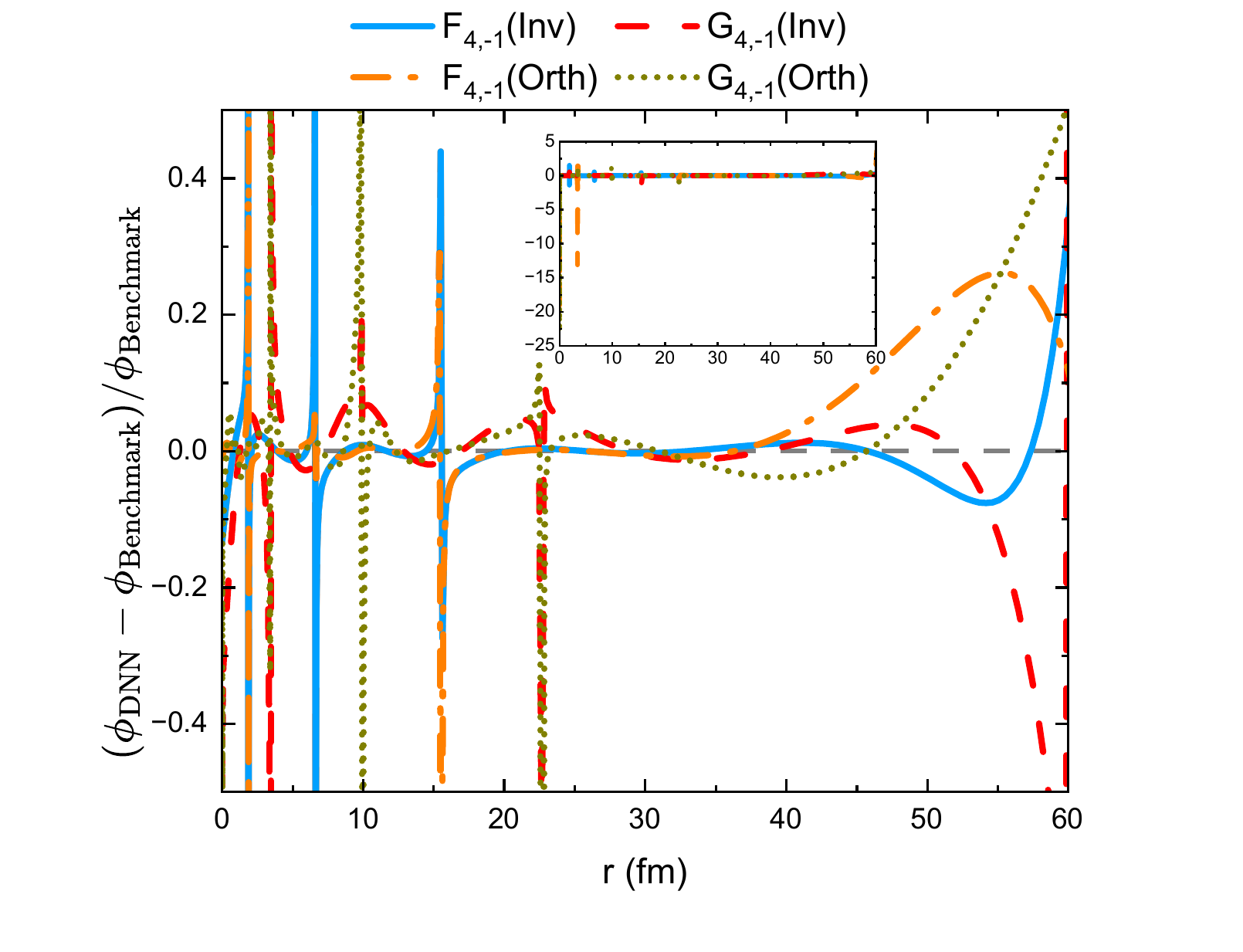}
  \end{minipage}
  \hfill
  \begin{minipage}{0.32\textwidth}
    \centering
    \includegraphics[width=1.0\linewidth]{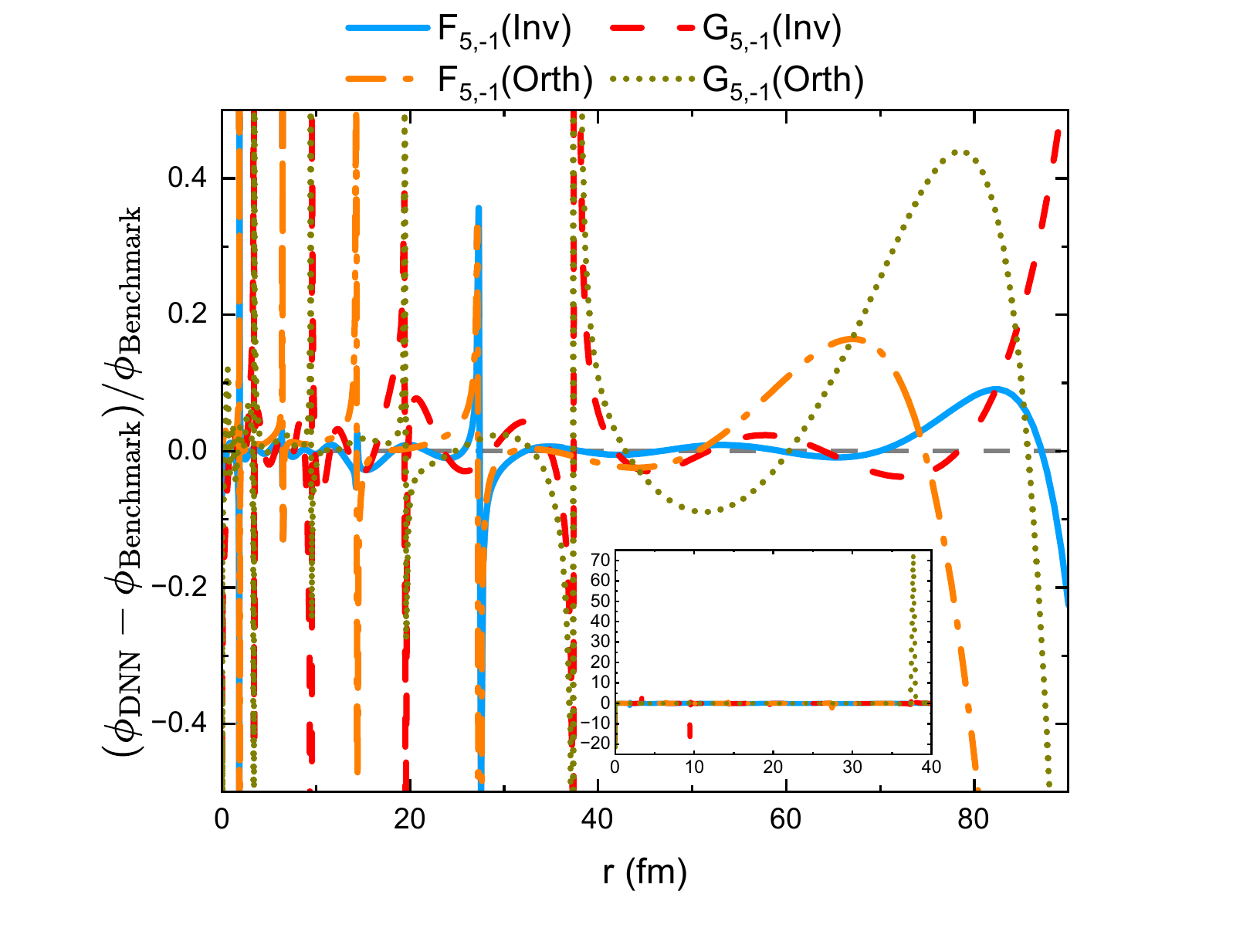}
  \end{minipage}
  \hfill
  \begin{minipage}{0.32\textwidth}
    \centering
    \includegraphics[width=1.0\linewidth]{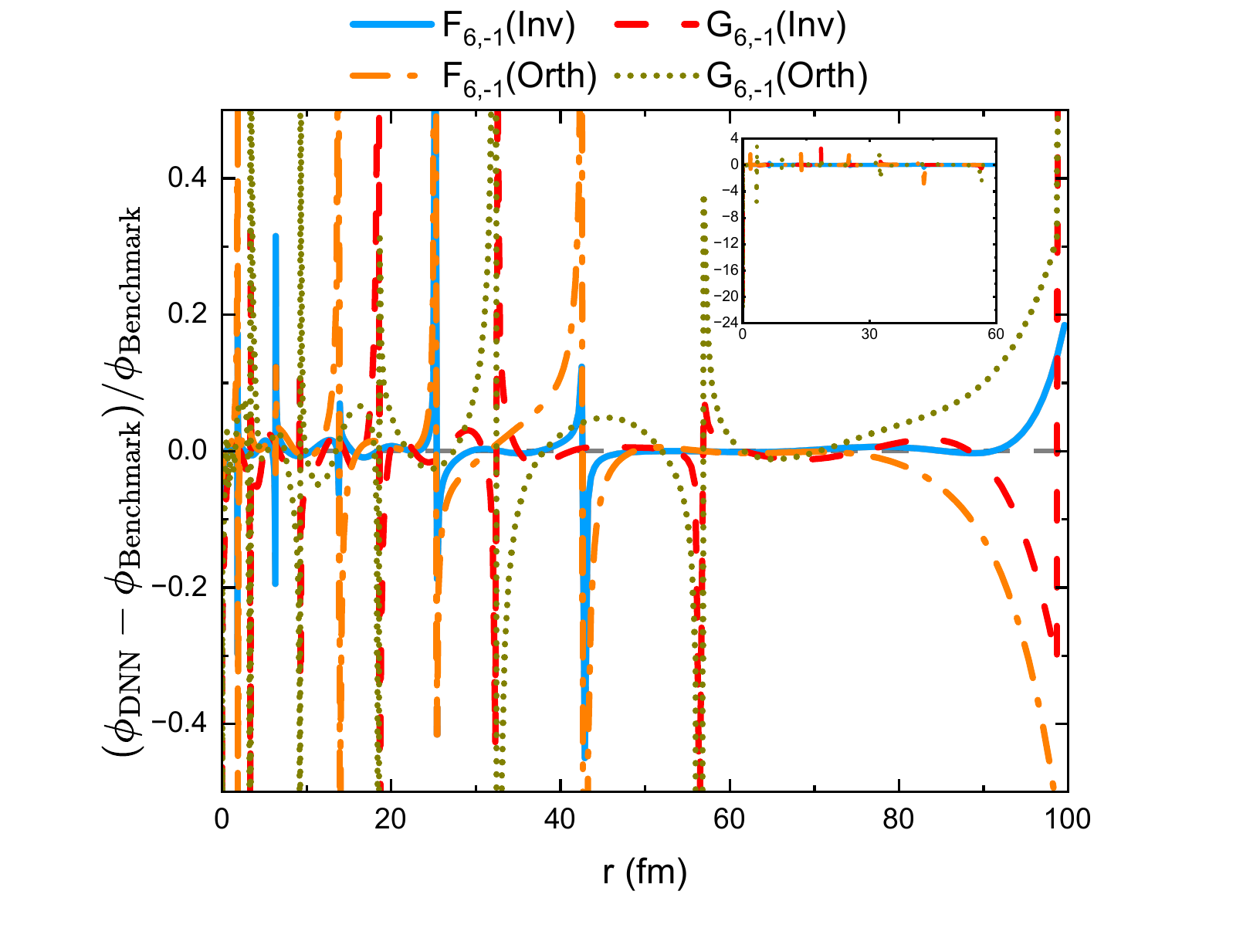}
  \end{minipage}
  \caption{
    Relative errors of DNN wave functions to the exact benchmark results for a hydrogen atom.
    The positions of peaks correspond to the node of wave functions.}
  \label{Coulomb_loss}
\end{figure*}

The wave functions of the ground and first five excited states for $ \kappa = -1 $ are shown in Figs.~\ref{F-1} and \ref{G-1}.
In general, both methods can reproduce the exact wave functions, except the peak of the $ n = 6 $ state at large $ r $, since the $ 100 \, \mathrm{a.u.} $ box size is too small for the $ n = 6 $ state.
The relative errors between DNN wave functions and exact results are shown in Fig.~\ref{Coulomb_loss}.
Except around the boundary, the DNN reproduces the exact wave functions.
The precisions of wave functions are almost the same level for both methods.
Note that due to the existence of nodes, the relative error diverges at those points.
\par
\begin{figure*}[tb]
  \begin{minipage}{0.48\textwidth}
    \centering
    \includegraphics[width=1.0\linewidth]{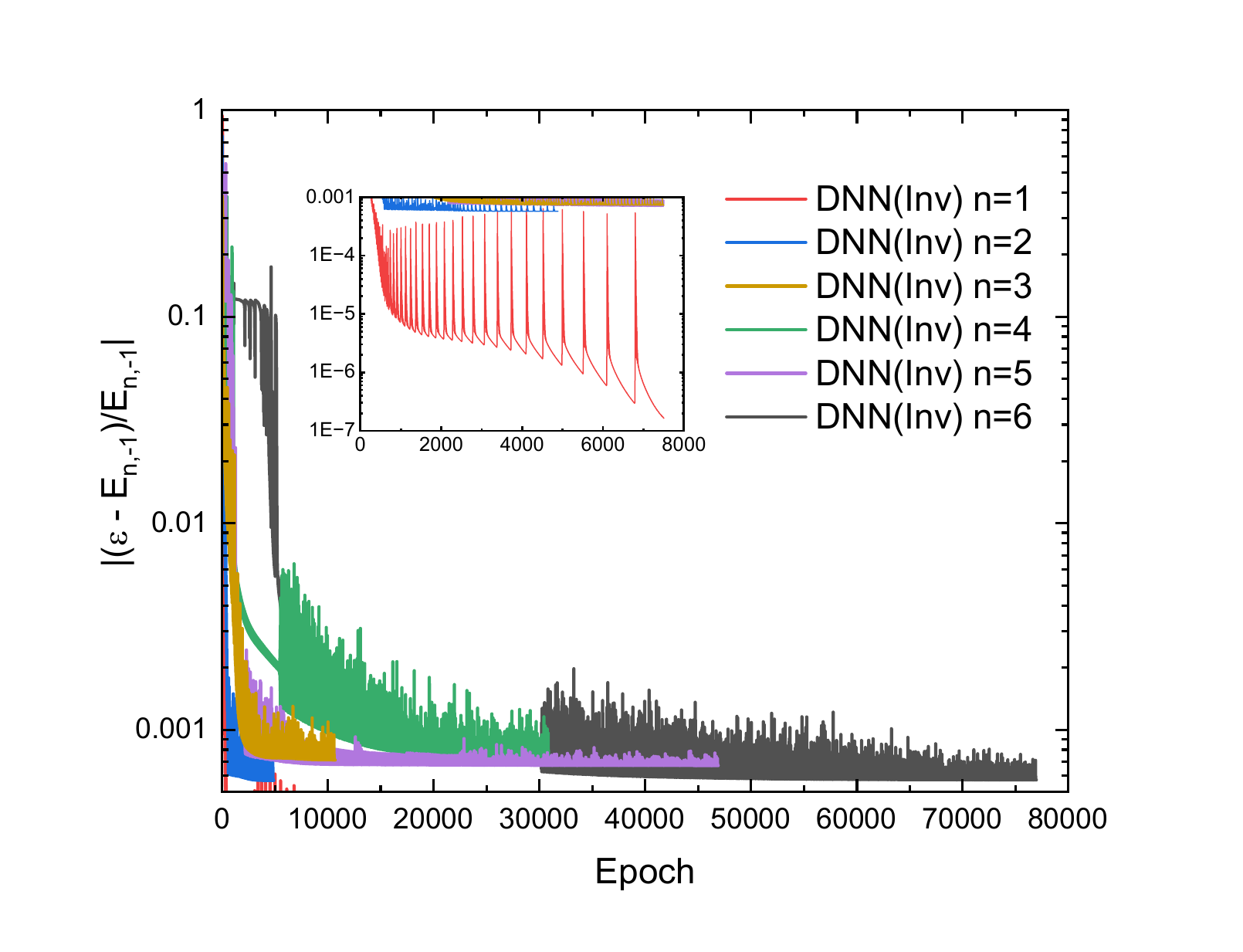}
  \end{minipage}
  \hfill
  \begin{minipage}{0.48\textwidth}
    \centering
    \includegraphics[width=1.0\linewidth]{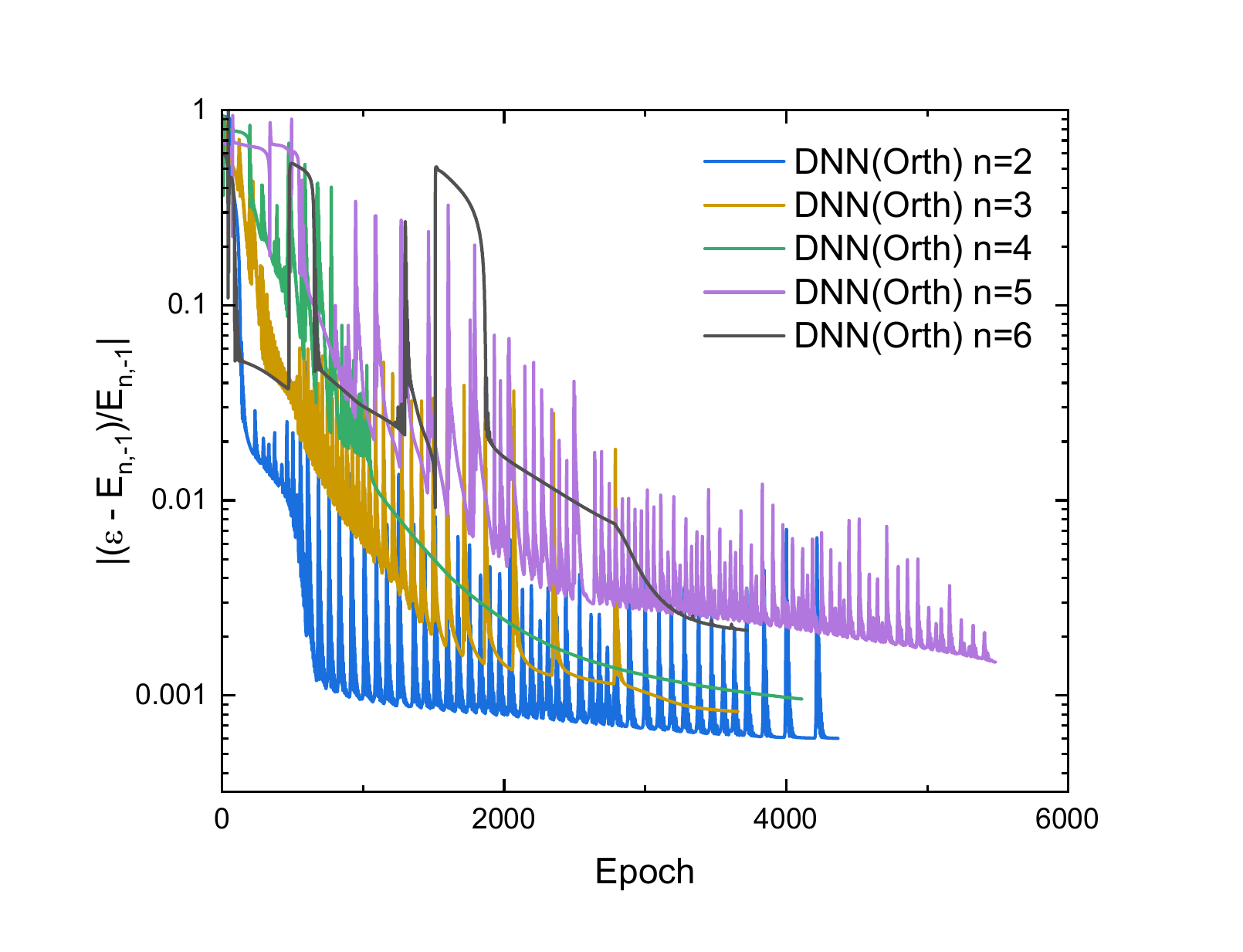}
  \end{minipage}
  \caption{
    Relative errors of energy of different states of $ \kappa = -1 $ as functions of epochs for a hydrogen atom.
    Because of the randomness in the optimization process, the number of epochs fluctuates in each run.
    The left panel is for the inverse Hamiltonian method, and the right panel is for the orthonormal method.}
  \label{closs}
\end{figure*}

The relative errors of the loss function to the exact energy as a function of epochs for both methods are shown in Fig.~\ref{closs}.
As shown in Table~\ref{Hkappa-1}, the number of epochs to converge for the orthonormal method is one order of magnitude less than the inverse Hamiltonian method, while the time per epoch of the orthonormal method is about twice longer than the inverse Hamiltonian method.
The calculation cost of the inverse Hamiltonian method does not depend on $ n $;
in contrast, the longer calculation for the orthonormal method is required for larger $ n $ due to the orthonormalization.
\par
\begin{table*}[tb]
  \centering
  \caption{
    Energies with $ \kappa = -1$ of the hydrogen atom from $ n = 1 $ to $ 6 $.
    The $ \epsilon_{\urm{Inv}} $ and $ \epsilon_{\urm{Orth}} $ denote the energy obtained by the inverse Hamiltonian method and the orthonormal one, respectively.
    Rows with ``---'' in the column denote this state cannot be calculated by the orthonormal method.
    Except for the $ n = 5 $ and $ n = 6 $ states with the orthonormal method achieving the $ 1 \times 10^{-3} $ level of precision, the rest calculations reach at least $ 1 \times 10^{-4} $ level of precision.
  }
  \label{Hkappa-1}
  \begin{ruledtabular}
    \begin{tabular}{lcccccc}
      $ n $ & 1 & 2 & 3 & 4  & 5 & 6 \\
      \hline
      Exact (Hartree) & $ -0.50000666 $ & $ -0.125002 $ & $ -0.055556 $ & $ -0.0312503 $ & $ -0.020000 $ & $ -0.013889 $ \\
      \hline
      $ \epsilon_{\urm{Inv}}$ (Hartree) & $ -0.50000657 $ & $ -0.124931 $ & $ -0.055516 $ & $ -0.0312271 $ & $ -0.019987 $ & $ -0.013881 $ \\
      Relative Error of $ \epsilon_{\urm{Inv}} $ & $ 1.79 \times 10^{-7} $ & $ 5.68 \times 10^{-4} $ & $ 7.20 \times 10^{-4} $ & $ 7.42 \times 10^{-4} $ & $ 6.50 \times 10^{-4} $ & $ 5.76 \times 10^{-4} $ \\
      Time per Epoch of $ \epsilon_{\urm{Inv}} $ ($ \mathrm{s} $) & $ 0.0553 $ & $ 0.0562 $ & $ 0.0575 $ & $ 0.0567 $ & $ 0.0557 $ & $ 0.0570 $ \\
      $ \epsilon'_{n -1} $ (Hartree) & $ -0.51 $ & $ -0.13 $ & $ -0.06 $ & $ -0.04 $ & $ -0.021 $ & $ -0.015 $ \\
      \hline
      $ \epsilon_{\urm{Orth}} $ (Hartree) & --- & $ -0.124927 $ & $ -0.055510 $ & $ -0.0312204 $ & $ -0.019970 $ & $ -0.013859 $ \\
      Relative Error of $ \epsilon_{\urm{Orth}} $ & --- & $ 6.00 \times 10^{-4} $ & $ 8.28 \times 10^{-4} $ & $ 9.57 \times 10^{-4} $ & $ 1.50 \times 10^{-3} $ & $ 2.16 \times 10^{-3} $\\
      Time per Epoch of $ \epsilon_{\urm{Orth}} $ ($ \mathrm{s} $) & --- & $ 0.111 $ & $ 0.120 $ & $ 0.125 $ & $ 0.126 $ & $ 0.136 $
    \end{tabular}
  \end{ruledtabular}
\end{table*}
\clearpage
%
\section{Results for Woods-Saxon potentials}
\label{Sect:V}
%
%
\subsection{The Woods-Saxon potentials}
\par
In this section, we apply the DNN model to a neutron in the Woods-Saxon potentials of $ \nuc{O}{16}{} $ and $ \nuc{Pb}{208}{} $ nuclei.
In this section, we use $\hbar = c = 1$.
\par
The scalar potential $ S \left( r \right) $ and vector one $ V \left( r \right) $ of Woods-Saxon potentials for neutrons, respectively, read
\begin{subequations}
  \begin{align}
    \label{eq42}
    U \left( r \right) 
    & =
      V \left( r \right) + S \left( r \right) 
      =
      \frac{V_N^0}{1 + \exp \left( \left( r - R_0 \right) / a \right)},\\
    W \left( r \right) 
    & =
      V \left( r \right) - S \left( r \right) 
      = 
      \frac{- \lambda_n V_N^0}{1 + \exp \left( \left( r - R_0^{\urm{ls}} \right) / a^{\urm{ls}} \right)},\\
    V_N^0 
    & =
      V' \left( 1 - \kappa' \frac{N - Z}{N + Z} \right),\\
    R_0 
    & =
      r_0
      A^{\frac{1}{3}},\\
    R_0^{\urm{ls}} 
    & =
      r_0^{\urm{ls}}
      A^{\frac{1}{3}},
  \end{align}
\end{subequations}
with $ \lambda_n = 11.12 $,
$ V' = -71.28 \, \mathrm{fm} $,
$ \kappa' = 0.462 $,
$ r_0 = 1.233 \, \mathrm{fm} $,
$ a = 0.615 \, \mathrm{fm} $,
$ r_0^{\urm{ls}} = 1.144 \, \mathrm{fm} $
and $ a^{\urm{ls}} = 0.648 \, \mathrm{fm} $~\cite{koepf1991spin},
where $ A $, $ N $, and $ Z $ denote the number of nucleons, neutrons and protons, respectively. 
%
%
\subsection{Discrete representations with uniform mesh}
\label{Sect:V,B}
\par
The uniform mesh is adopted to calculate for the Woods-Saxon potential,
where $ r $ is uniformly distributed with mesh size $ \Delta r $.
The matrix of the first-order differential operator $ \frac{\partial}{\partial r} $ reads
\begin{equation}
  \frac{\partial}{\partial r}
  \simeq
  \frac{1}{2 \Delta r}  
  \begin{pmatrix}
    0       & 1      & 0      & \cdots & 0      & 0      & 0 \\
    -1      & 0      & 1      & \cdots & 0      & 0      & 0 \\
    0       & -1     & 0      & \cdots & 0      & 0      & 0 \\
    \vdots  & \vdots & \vdots & \ddots & \vdots & \vdots & \vdots \\
    0       & 0      & 0      & \cdots & 0      & 1      & 0 \\
    0       & 0      & 0      & \cdots & -1     & 0      & 1 \\
    0       & 0      & 0      & \cdots & 0      & -1     & 0
  \end{pmatrix}.
\end{equation}
%
%
\subsection{DNN results}
\par
In this section, we use a $ 20 \, \mathrm{fm} $ box with $ M = 2000 $ meshes for the DNN calculation.
The general pseudospectral method (GPS)~\cite{
  yao1993generalized,PhysRevA.104.022801}
is adopted to calculate benchmark wave functions and energies with the mapping parameter $ L = 10 $ and the number of mesh points $ 600 $.
\par
\begin{figure*}[tb]
  \begin{minipage}{0.48\textwidth}
    \centering
    \includegraphics[width=1.0\linewidth]{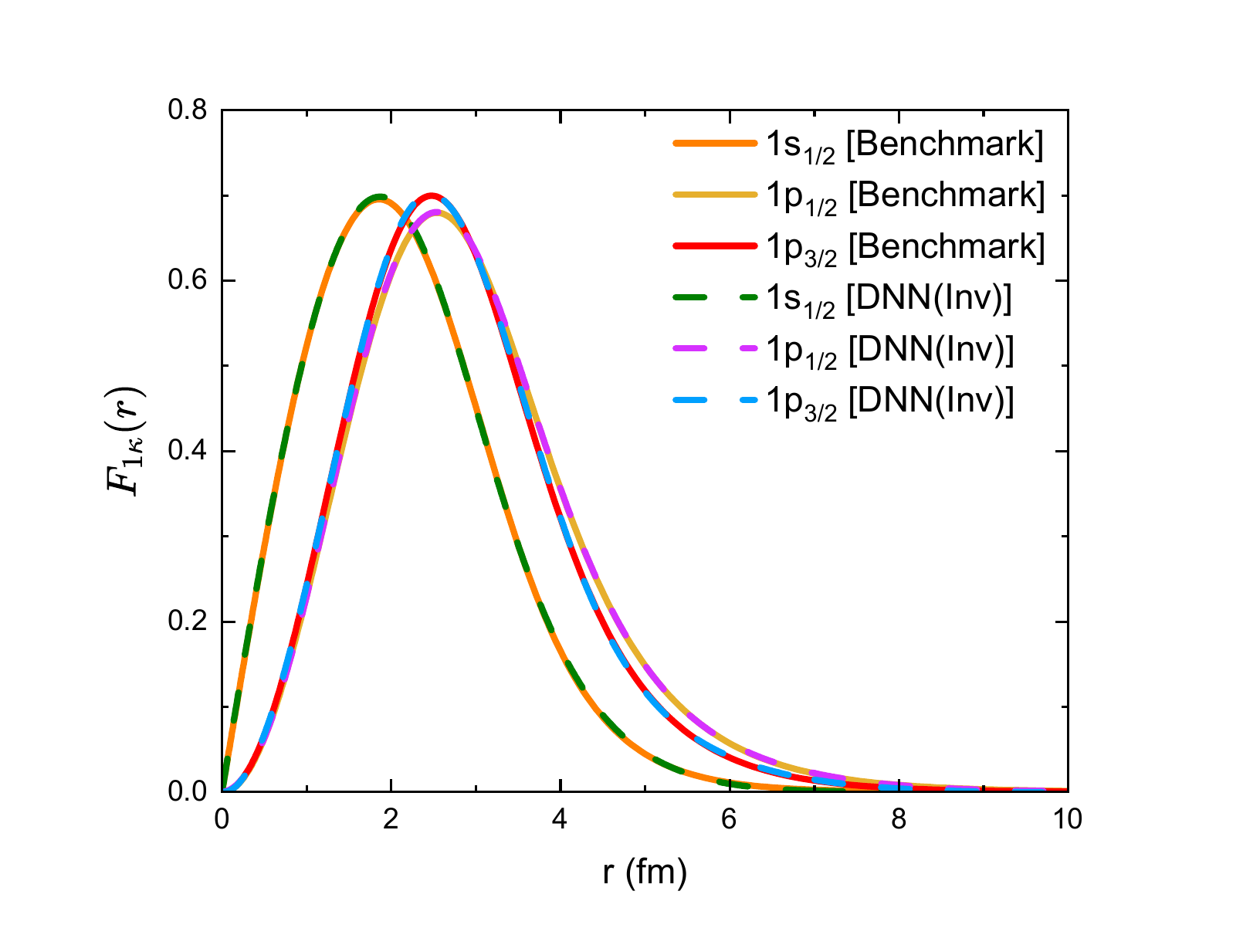}
  \end{minipage}
  \hfill
  \begin{minipage}{0.48\textwidth}
    \centering
    \includegraphics[width=1.0\linewidth]{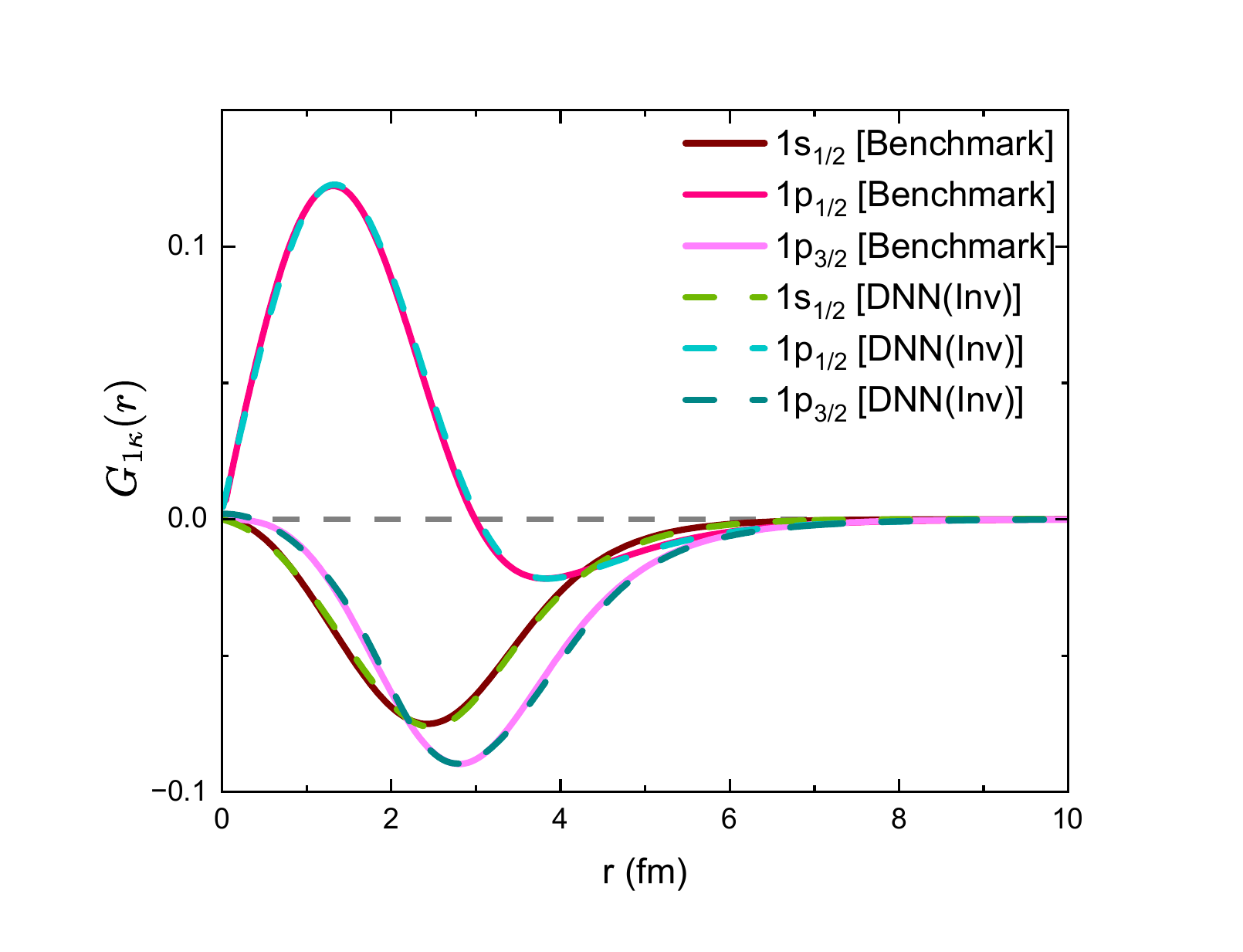}
  \end{minipage}
  \caption{
    Same as Figs~\ref{F-1} and \ref{G-1} but for the Woods-Saxon potential for $ \nuc{O}{16}{} $ below the Fermi energy.}
  \label{16Owave}
\end{figure*}

Table~\ref{16Otable} lists the energies of three states below the Fermi energy for
$ \nuc{O}{16}{} $ ($ Z = N = 8 $)
by the inverse Hamiltonian method, i.e., the $ 1s_{1/2} $, $ 1p_{3/2} $, and $ 1p_{1/2} $ states.
All the states are ``ground state'' for the given $ l $ and $ j $ (i.e., $ \kappa $).
The precision reaches at least $ 1 \times 10^{-4} $.
Figure \ref{16Owave} shows the corresponding wave functions.
The DNN wave functions show good agreements with the benchmark.
\par
\begin{figure}[tb]
  \centering
  \includegraphics[width=1.0\linewidth]{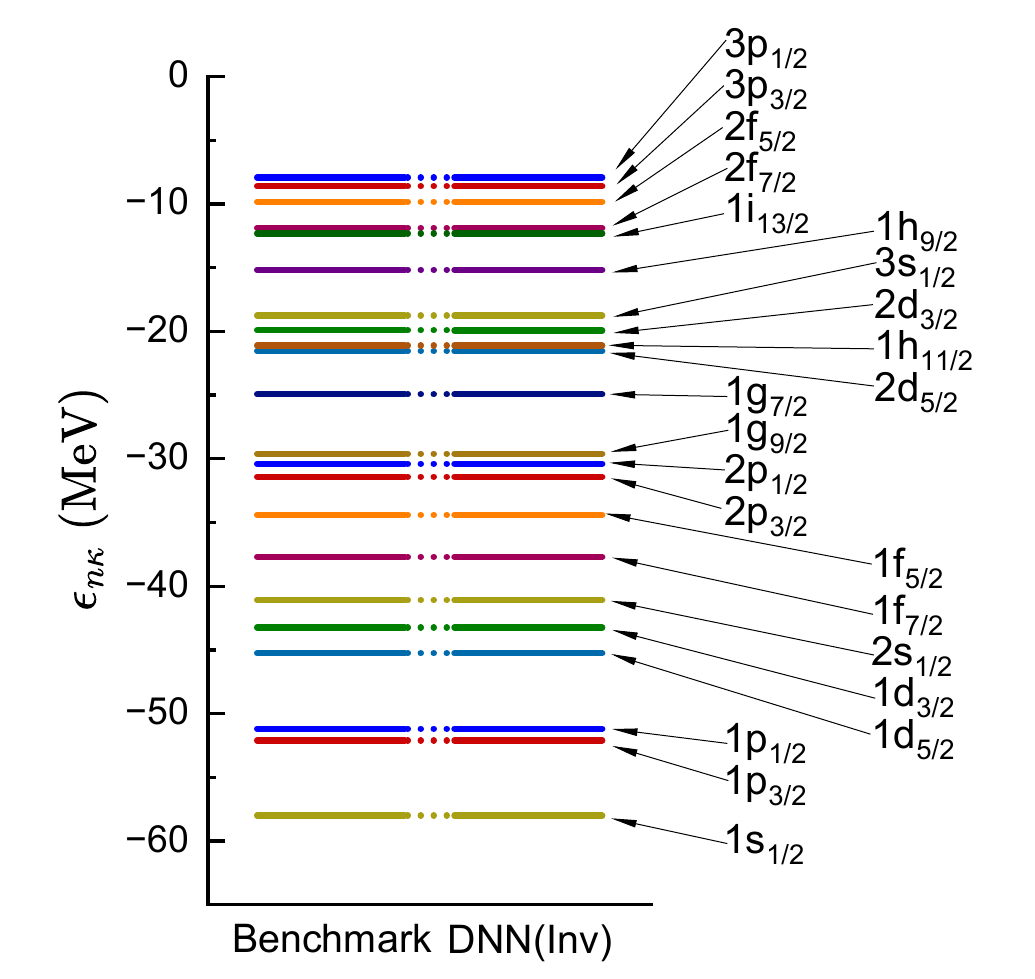}
  \caption{
    Spectrum of neutrons calculated by the Woods-Saxon potential of $ \nuc{Pb}{208}{} $.
    The benchmark is given by the general pseudospectral method and the inverse Hamiltonian method is used for the DNN.}
  \label{Pbspectrum}
\end{figure}

Figure~\ref{Pbspectrum} shows the energy spectrum of
$ \nuc{Pb}{208}{} $ ($ Z = 82 $, $ N = 126 $) below the Fermi energy.
The left-hand side is the benchmark and the right-hand side is the DNN results with the inverse Hamiltonian method.
All the energies obtained by the DNN are identical (within $ 0.002 \, \% $) to the benchmark calculation.
The first three states of $ \kappa = -1 $ as an example are listed in Table~\ref{Pb1} with both the inverse Hamiltonian and the orthonormal method, which give almost the same accuracy.
\par
\begin{figure*}[tb]
  \begin{minipage}{0.32\textwidth}
    \centering
    \includegraphics[width=1.0\linewidth]{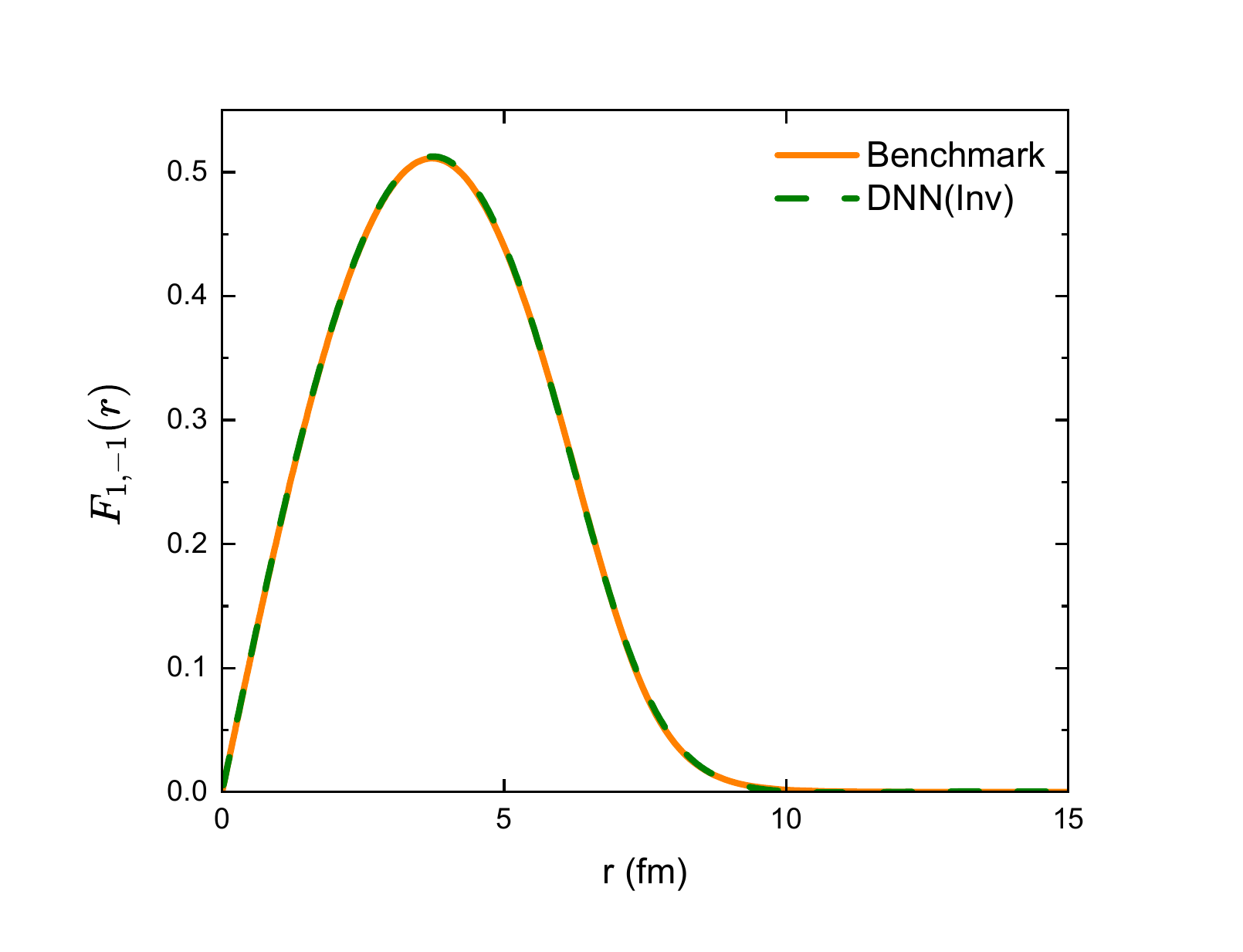}
  \end{minipage}
  \hfill
  \begin{minipage}{0.32\textwidth}
    \centering
    \includegraphics[width=1.0\linewidth]{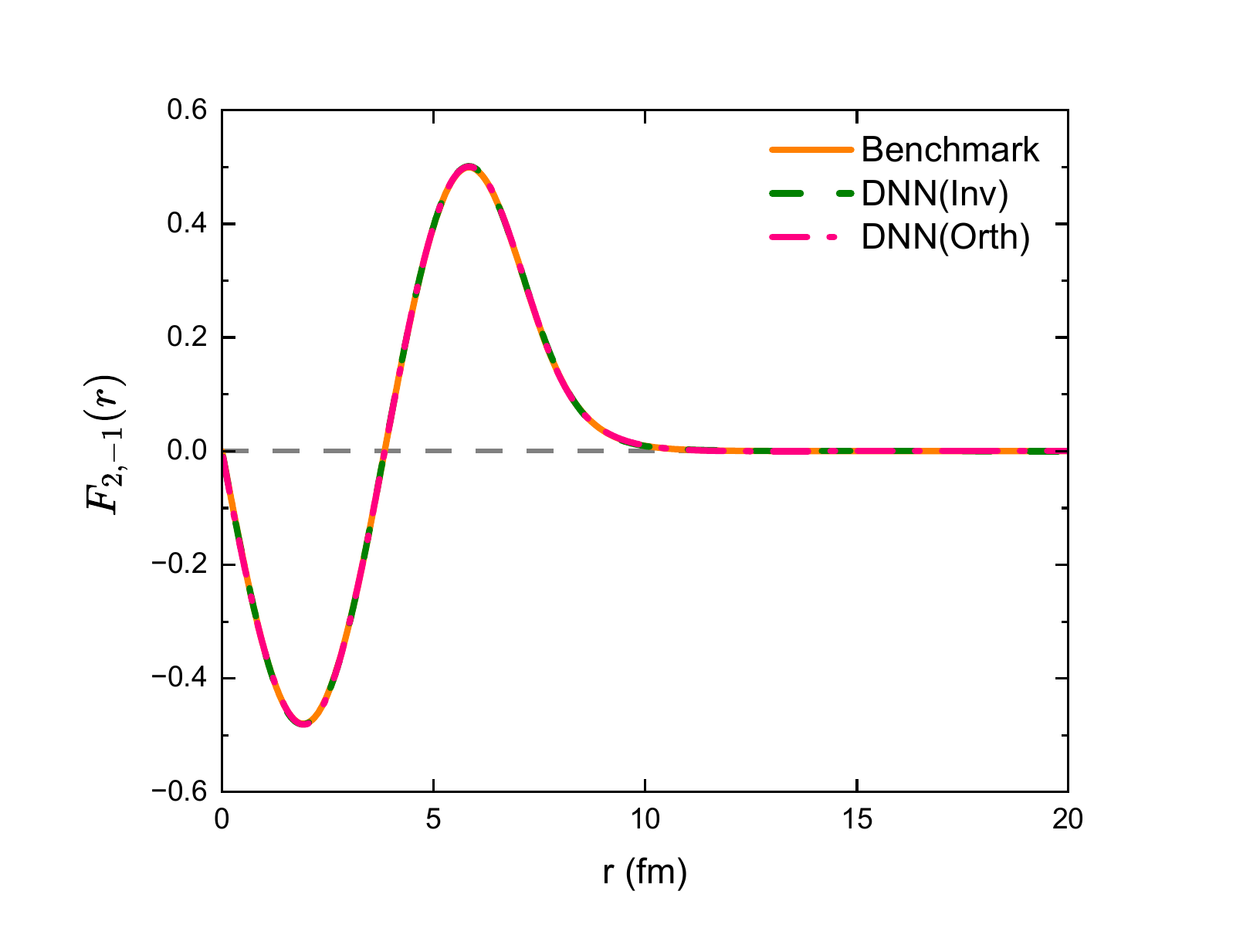}
  \end{minipage}
  \hfill
  \begin{minipage}{0.32\textwidth}
    \centering
    \includegraphics[width=1.0\linewidth]{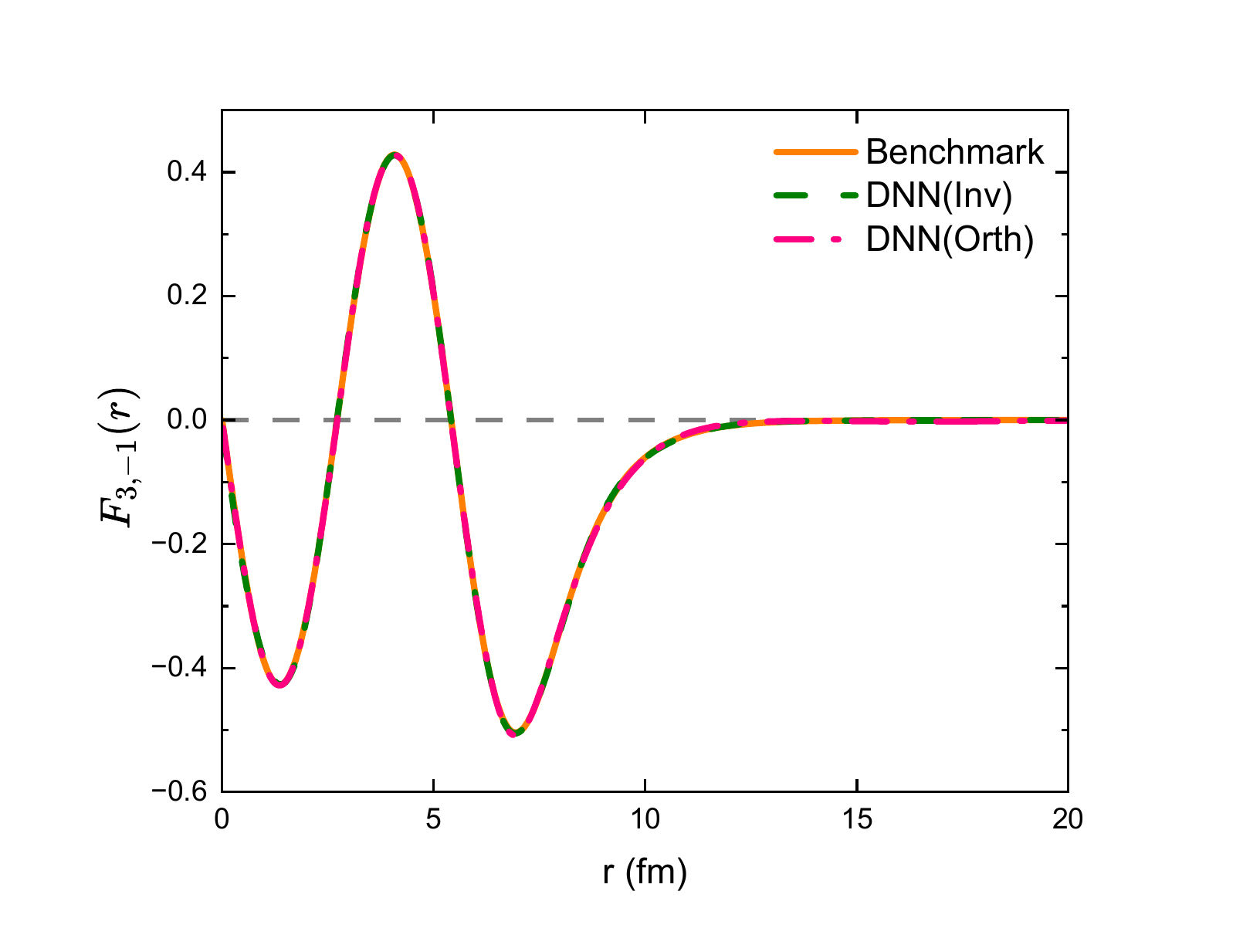}
  \end{minipage}
  \caption{
    Same as Fig.~\ref{F-1} but for the Woods-Saxon potential for $ \nuc{Pb}{208}{} $.
    Only $\kappa = -1$ states below the Fermi energy are shown.}
  \label{PbF1}
\end{figure*}
\begin{figure*}[tb]
  \begin{minipage}{0.32\textwidth}
    \centering
    \includegraphics[width=1.0\linewidth]{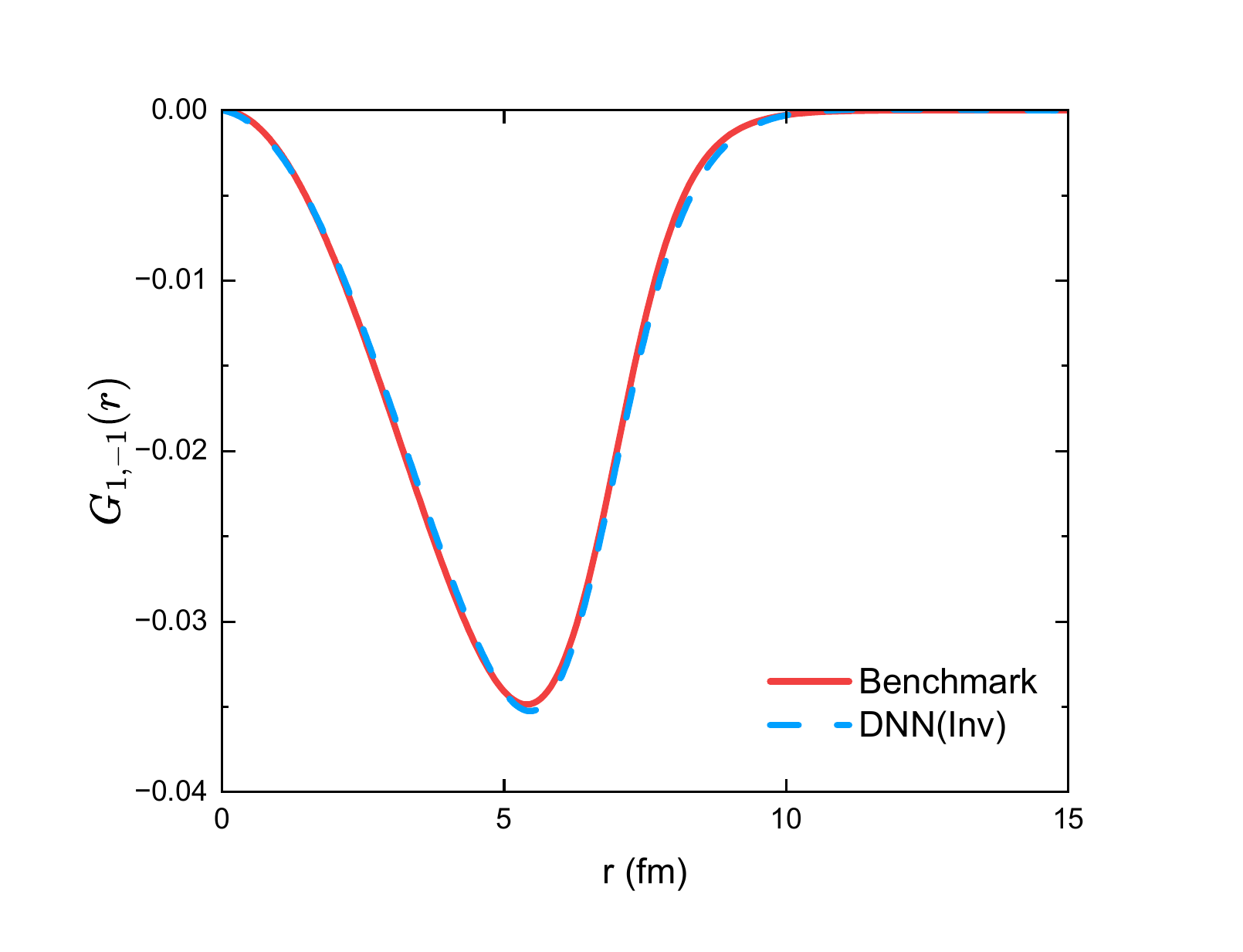}
  \end{minipage}
  \hfill
  \begin{minipage}{0.32\textwidth}
    \centering
    \includegraphics[width=1.0\linewidth]{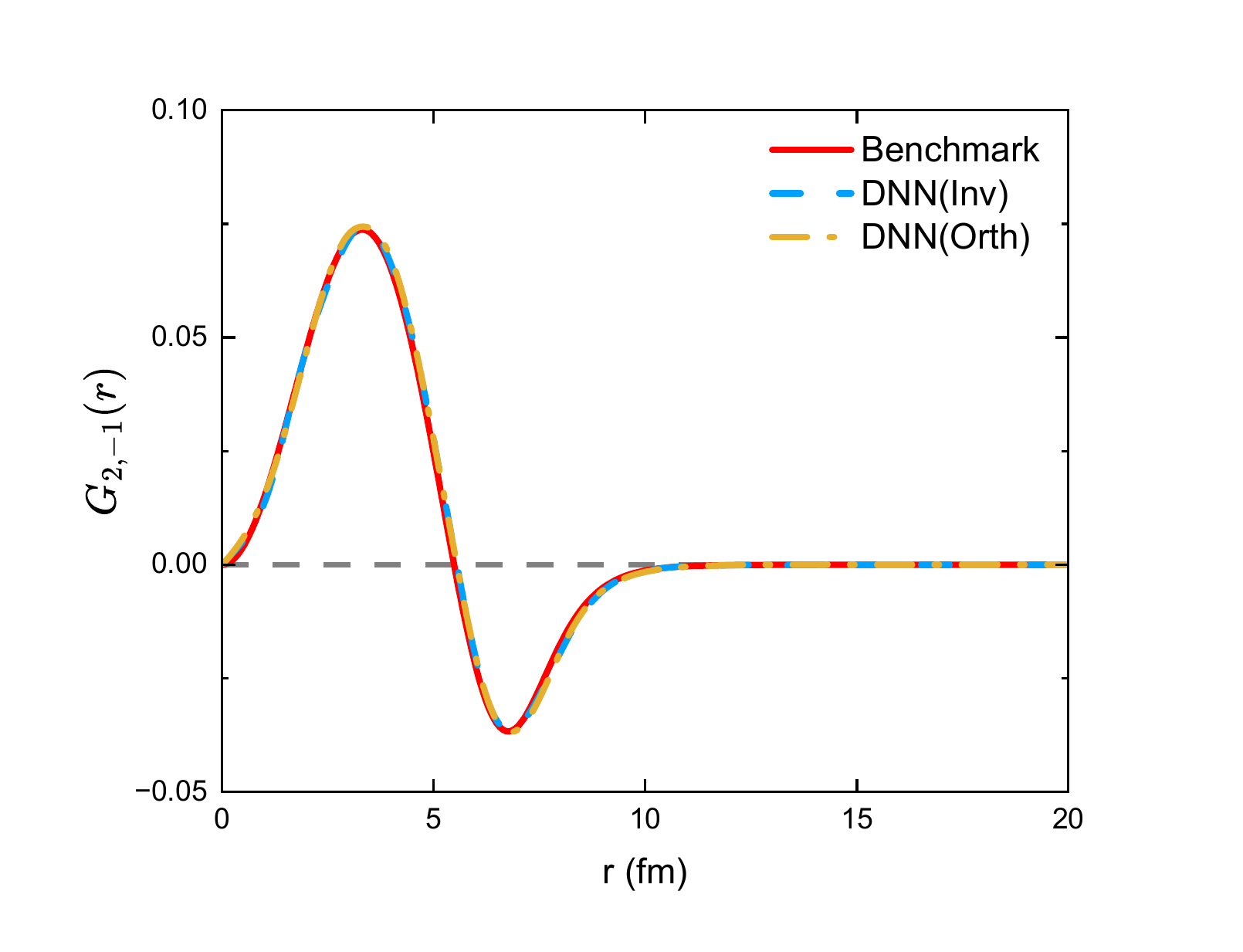}
  \end{minipage}
  \hfill
  \begin{minipage}{0.32\textwidth}
    \centering
    \includegraphics[width=1.0\linewidth]{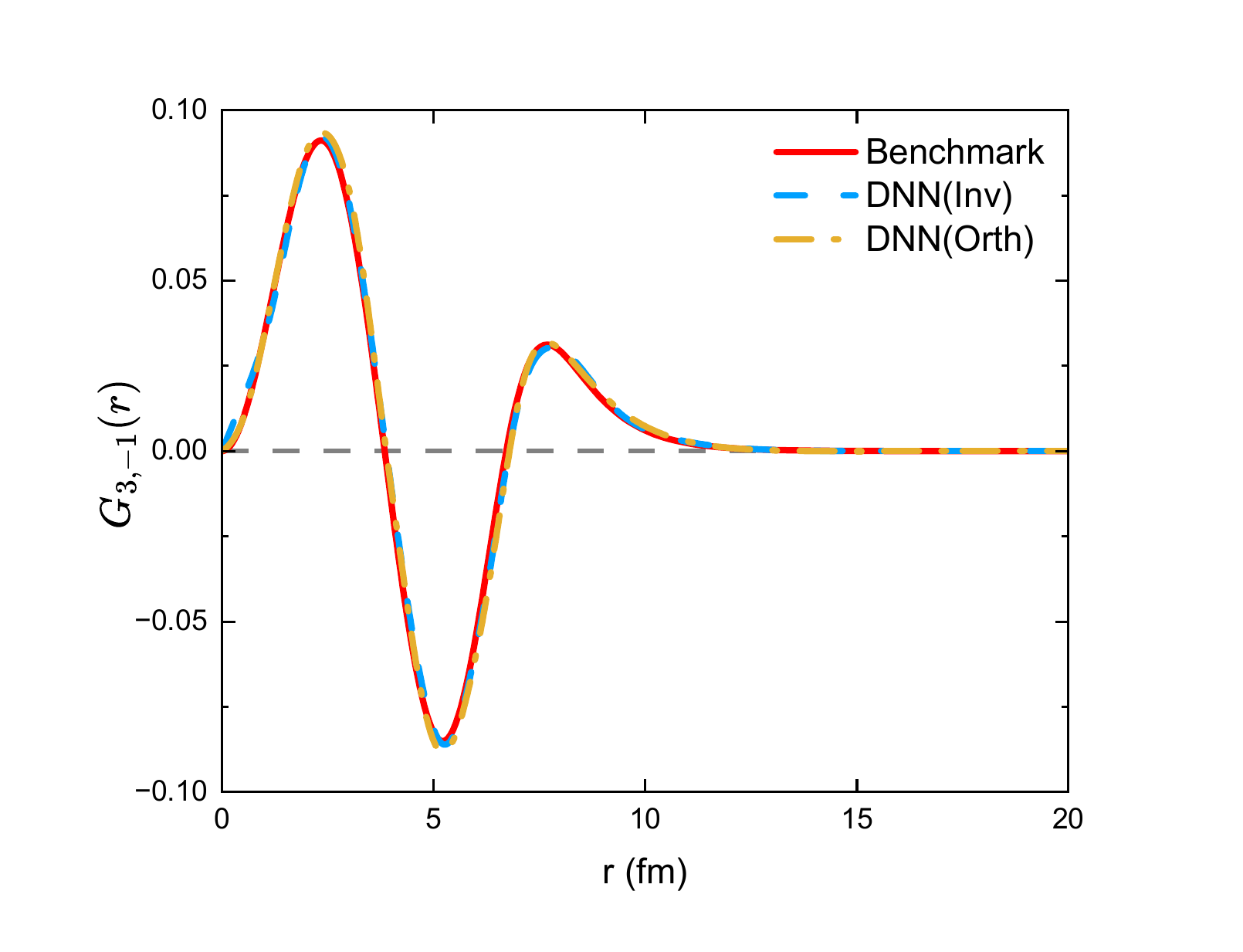}
  \end{minipage}
  \caption{
    Same as Fig.~\ref{PbF1} but for the $ G $-component.}
  \label{PbG1}
\end{figure*}

The $ F $- and $ G $-components for $ \kappa = -1 $ of $ \nuc{Pb}{208}{} $ are shown in Figs.~\ref{PbF1} and \ref{PbG1}, respectively.
Both methods show good agreements with the benchmark for the $ F $-component, while the inverse Hamiltonian method gives slightly more accurate results than the orthonormal method for the $ G $-component of $ n = 3 $ state.
According to Eq.~\eqref{eq31}, the accuracy of the orthonormal method depends on the accuracy of lower-state wave functions, which means it also depends on the inverse Hamiltonian method which provides the lowest state data.
\par
For the GPS method, which solves the eigenvalue problem of a Dirac Hamiltonian, the computational cost scales as
$ O \left( M^3 \right) $, where $ M $ is the number of mesh points.
For our DNN framework with the inverse Hamiltonian method, the cost is approximately $ O \left( XM^2 \right) $,
where $ X $ is the number of epochs.
As shown in Fig.~\ref{closs}, $ X > M $ holds in our example.
However, we can use an easier convergence condition to stop the DNN program at a smaller $ X $ to reduce the computational cost.
Additionally, $ X $ remains nearly constant with respect to $ M $, which suggests that the DNN method may become more computationally favorable for larger systems requiring larger mesh sizes.
\begin{table}[tb]
  \centering
  \caption{
    Energies of $ \nuc{O}{16}{} $ below the Fermi energy calculated by the Woods-Saxon potential.
    The benchmark is obtained by the general pseudospectral method.}
  \label{16Otable}
  \begin{ruledtabular}
    \begin{tabular}{lccc}
      $ \kappa $                                  & $ -1 $                  & $ -2 $                  & $ 1 $        \\
      \hline
      Benchmark ($ \mathrm{MeV} $)                & $ -43.16880 $           & $ -24.6354 $            & $ -18.9746 $ \\
      \hline
      $ \epsilon_{\urm{Inv}} $ ($ \mathrm{MeV} $) & $ -43.16859 $           & $ -24.6353 $            & $ -18.9772 $ \\
      $ \epsilon'_{n \kappa} $ ($ \mathrm{MeV} $) & $ -45.0     $           & $ -28.0    $            & $ -20.0    $ \\
      Relative Error of $ \epsilon_{\urm{Inv}} $  & $ 4.86 \times 10^{-6} $ & $ 4.06 \times 10^{-6} $ & $ 1.37 \times 10^{-4} $
    \end{tabular}
  \end{ruledtabular}
\end{table}
\begin{table}[tb]
  \centering
  \caption{
    Same as Table \ref{16Otable} but for $ \nuc{Pb}{208}{} $.
    Only $ \kappa = -1 $ states are shown.}
  \label{Pb1}
  \begin{ruledtabular}
    \begin{tabular}{lccc}
      $n$                                          & $ 1 $                   & $ 2 $                   & $ 3 $\\
      \hline
      Benchmark ($ \mathrm{MeV} $)                 & $ -58.0026 $            & $ -41.0773 $            & $ -18.7560 $ \\
      \hline
      $ \epsilon_{\urm{Inv}} $ ($ \mathrm{MeV} $)  & $ -58.0028 $            & $ -41.0791 $            & $ -18.7626 $ \\
      $ \epsilon'_{n -1} $ ($ \mathrm{MeV} $)       & $ -60.0    $            & $ -45.0    $            & $ -20.0   $ \\
      Relative Error of $\epsilon_{\urm{Inv}}$     & $ 3.45 \times 10^{-6} $ & $ 4.38 \times 10^{-5} $ & $ 3.52 \times 10^{-4} $ \\
      \hline
      $ \epsilon_{\urm{Orth}} $ ($ \mathrm{MeV} $) & ---                     & $ -41.0789 $            & $ -18.7597 $\\
      Relative Error of $ \epsilon_{\urm{Orth}} $  & ---                     & $ 3.90 \times 10^{-5} $ & $ 1.97 \times 10^{-4} $
    \end{tabular}
  \end{ruledtabular}
\end{table}
%
%
\section{Summary}
\label{Sect:VI}
\par
In this paper, we extended the deep neural network method proposed in Ref.~\cite{
  PhysRevResearch.5.033189} to the Dirac equation, where the existence of the Dirac sea leads to the variational collapse for the usual variational method.
To avoid such a problem, we applied the inverse Hamiltonian method, which reverses the order of the energy spectrum so that the target state is at the bottom, thus making the method applicable to both the ground and excited states.
Besides the inverse Hamiltonian method, we also extended the orthonormal method~\cite{
  PhysRevResearch.5.033189}
to calculate excited states.
By applying the orthonormal condition, a state that is orthonormal to all the lower states is obtained to calculate the loss function so that components of all the lower states are eliminated in the DNN results.
\par
To verify the feasibility of our method, we calculate a hydrogen atom and the Woods-Saxon potentials of $ \nuc{O}{16}{} $ and $ \nuc{Pb}{208}{} $.
We found that both methods show good agreements with the benchmark with about the same accuracy (up to $ 0.15 \, \% $ difference for the listed states) for the both systems.
Additionally, our DNN can consistently reach the desired state regardless of the choice of weight initializers.
\par
One potential extension of our method could involve applications to relativistic many-body systems.
For instance, the two-body Dirac equation~\cite{
  CRATER198357, crater1999two}---which describes relativistic two-particle systems---can,
in the center-of-momentum frame, be simplified to a partial differential equation structurally similar to the nonrelativistic Schr\"{o}dinger equation.
In this form, the total energy emerges as an eigenvalue, suggesting that our deep neural network (DNN) approach, with suitable modifications, might be adapted to solve for both ground and excited states in such systems.
Additionally, while beyond the scope of the current work, our methodology may hold promise for addressing challenging problems in nuclear physics, such as single-particle resonances in weakly bound nuclei or other systems where traditional computational methods face limitations.
These potential applications remain subjects for future investigation.
%
%
\begin{acknowledgments}
  The authors thank Mario Centelles, Koji Hashimoto, and Hisashi Naito for the fruitful discussion.
  C.W.~acknowledges the warm hospitality and computational resources of the RIKEN iTHEMS program.
  T.N.~acknowledges 
  the RIKEN Special Postdoctoral Researcher Program,
  the JSPS Grant-in-Aid for Research Activity Start-up under Grant No.~JP22K20372,
  the JSPS Grant-in-Aid for Transformative Research Areas (A) under Grant No.~JP23H04526 and JP25H01558,
  the JSPS Grant-in-Aid for Scientific Research (S) under Grant No.~JP25H00402,
  the JSPS Grant-in-Aid for Scientific Research (B) under Grant Nos.~JP23H01845, JP23K26538, and JP25K01003,
  the JSPS Grant-in-Aid for Scientific Research (C) under Grant No.~JP23K03426,
  the JSPS Grant-in-Aid for Early-Career Scientists under Grant No.~JP24K17057,
  and
  the JSPS Grant-in-Aid for JSPS Fellows under Grant No.~JP25KJ0405.
  J.L.~acknowledges
  the National Natural Science Foundation of China (Nos.~12475119 and 11675063), Scientific Research Project of Education Department of Jilin Province (No.~JJKH20241242KJ).
  H.L.~acknowledges the JSPS Grant-in-Aid for Scientific Research (S) under Grant No.~JP20H05648 and the RIKEN Pioneering Project: Evolution of Matter in the Universe.
  The numerical calculations were performed on cluster computers at the RIKEN iTHEMS program.
\end{acknowledgments}
%
\clearpage
\appendix
\section{Failed training set-ups}
\label{Appen:A}
\par
In this appendix, we discuss which DNN structure and output fail to generate the accurate $ G $-component wave function and results of minimizing the energy without the inverse Hamiltonian.
%
%
\subsection{Employing a non-fully connected neural network}
\label{Appendix:B}
\par
\begin{figure}[tb]
  \centering
  \includegraphics[width=1.0\linewidth]{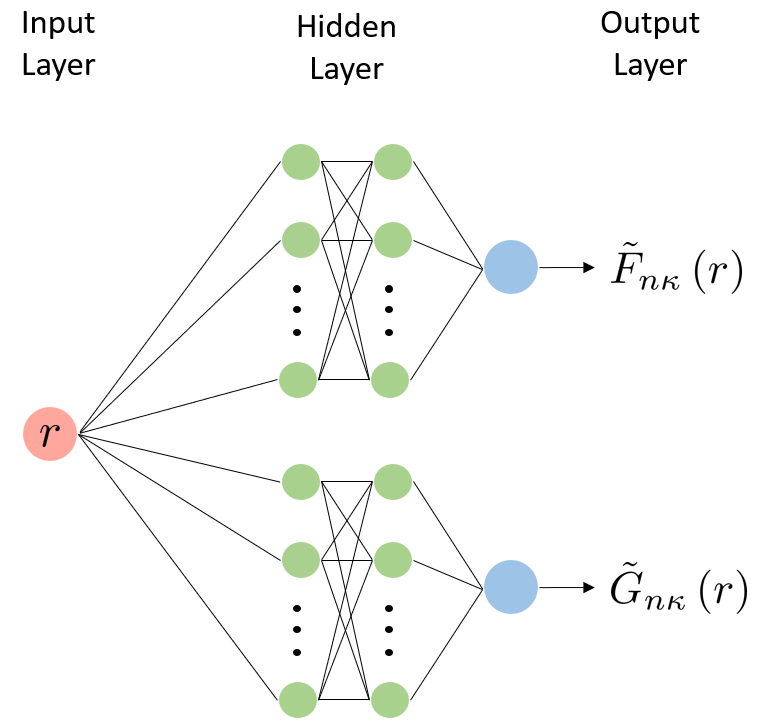} 
  \caption{The structure of the non-fully connected DNN. One output unit generates the $ F $-component of the wave function, while the other generates the $ G $-component.
    For each output, the units in the hidden layers are fully connected, while the hidden layer units of different outputs are not connected to each other.}
  \label{DNN2}
\end{figure}

Besides the fully-connected DNN used in the main text, we also test a non-fully connected deep neural network to generate the $ F $- and $ G $-components, whose structure is shown in Fig.~\ref{DNN2},
where $ F_{n \kappa} $ and $ G_{n \kappa} $ are generated by two output units, respectively.
For each output, the units of the hidden layers are fully connected, while the hidden layer units of different outputs are not connected to each other. 
\par
\begin{figure}[tb]
  \centering
  \includegraphics[width=1.0\linewidth]{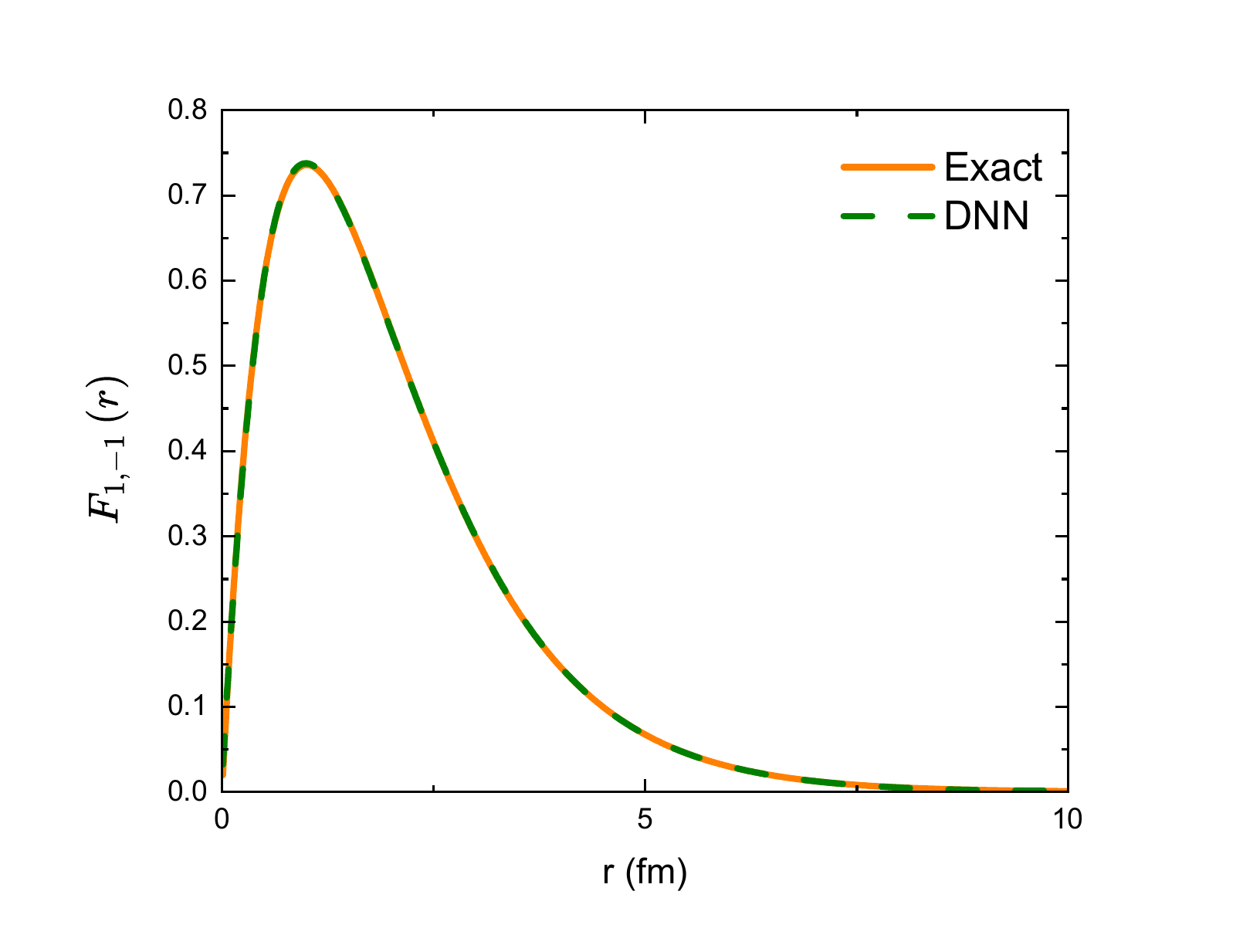} 
  \includegraphics[width=1.0\linewidth]{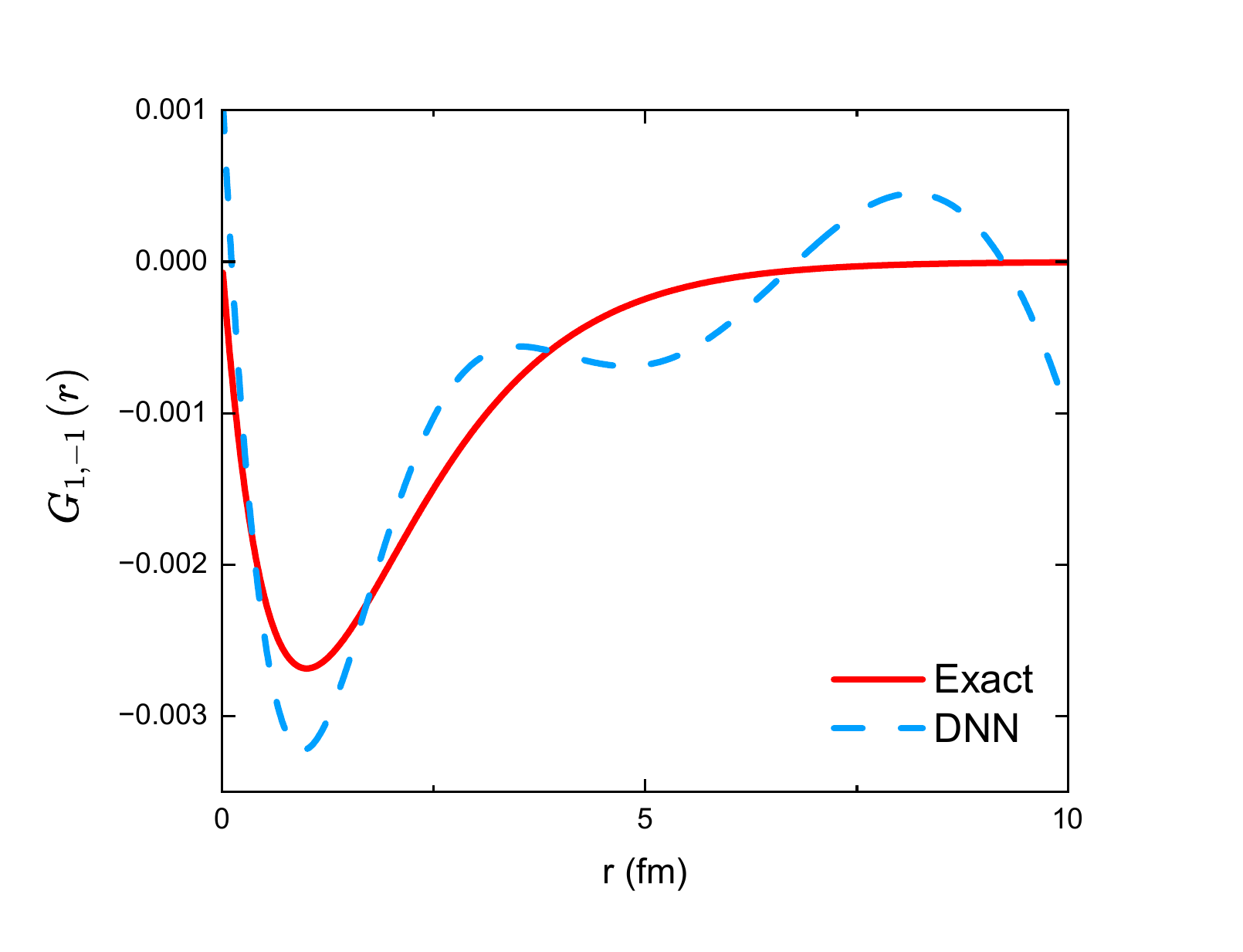} 
  \caption{The ground-state wave functions of a hydrogen atom calculated by the non-fully connected DNN.
    The $ F $-component is consistent with the benchmark, while the $ G $-component is largely different from the benchmark.}
  \label{DNN2Coulomb}
\end{figure}
\begin{figure}[tb]
  \centering
  \includegraphics[width=1.0\linewidth]{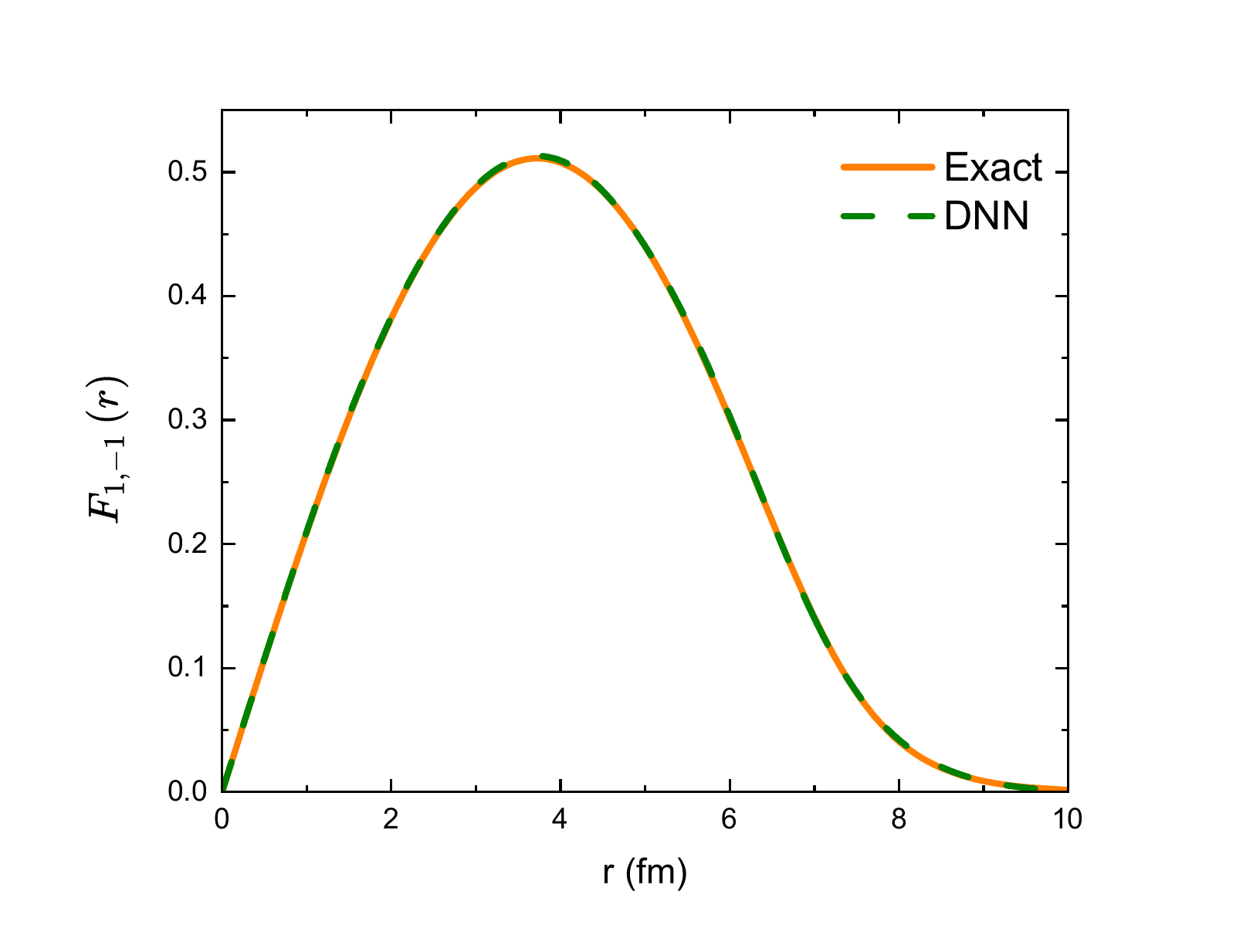} 
  \includegraphics[width=1.0\linewidth]{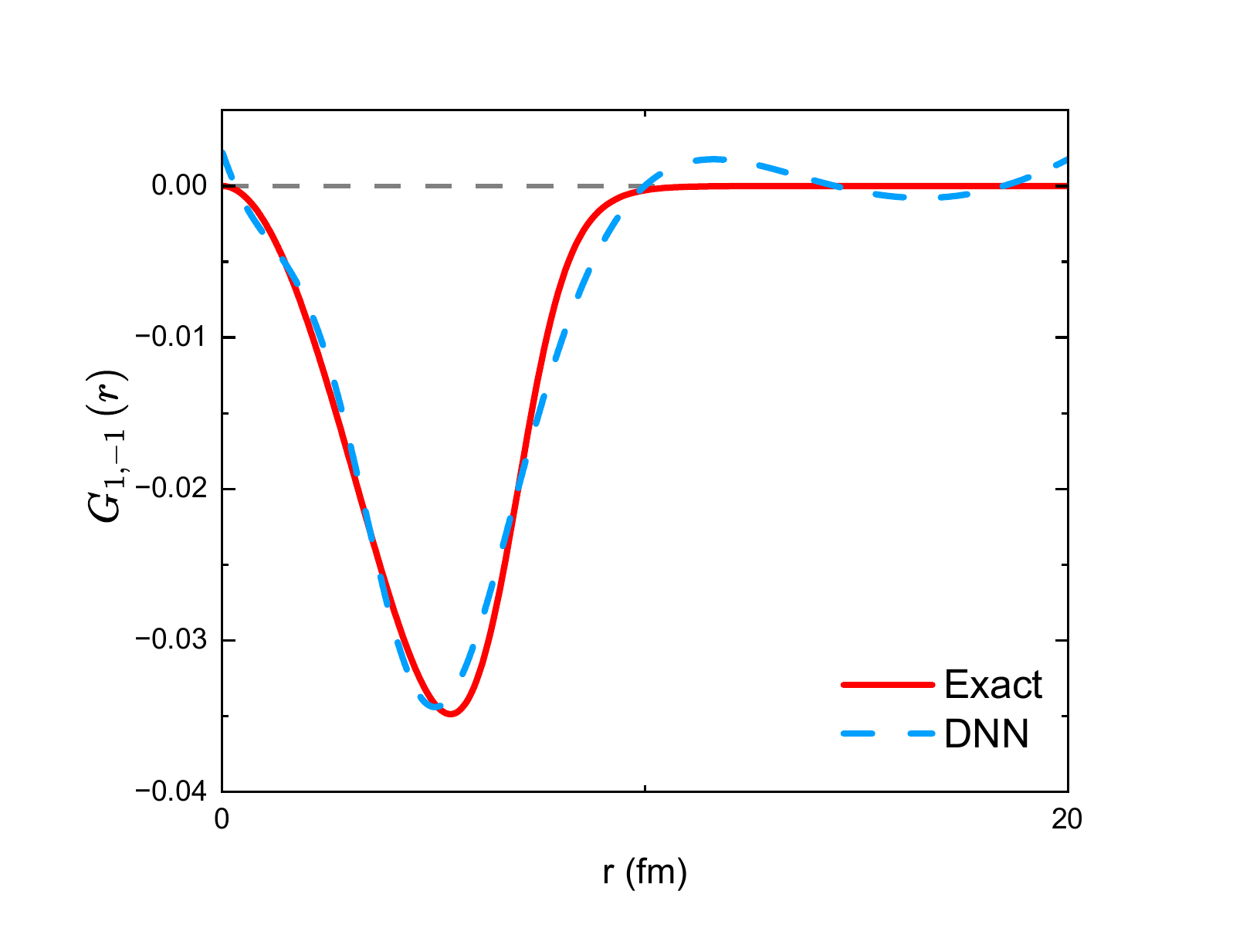} 
  \caption{
    Same as Fig.~\ref{DNN2Coulomb} but for the Woods-Saxon potential for $ \nuc{Pb}{208}{} $.}
  \label{DNN2WS}
\end{figure}

As shown in Fig.~\ref{DNN2Coulomb}, the $ G $-component is largely different from the benchmark.
This is because the $ G $-component contributes to the norm [Eq.~\eqref{eq12_}] in about $ 0.001 \, \% $,
and thus its contribution to the loss function is too small for optimization.
Therefore, even if the $ G $-component behaves strangely, the loss function can easily satisfy the convergence criteria.
Although the $ G $-component contributes to the norm in $ 0.39 \, \% $ in the case of $ \nuc{Pb}{208}{} $,
it is still not large enough to be well optimized with the non-fully connected neural network as shown in Fig.~\ref{DNN2WS}.
%
%
\subsection{Usage of the trial wave function $f_{n \kappa}$}
\label{Appendix:C}
\par 
In Sec. \ref{Sect:III,A}, a trial wave function $ f_{n \kappa} $ is used as the DNN output to calculate $ F_{n \kappa} $ and $ G_{n \kappa} $.
The denominator $ r $ improves the accuracy of $ F_{n \kappa} $ and $ G_{n \kappa} $ close to the origin.
\par
\begin{figure}[tb]
  \centering
  \includegraphics[width=1.0\linewidth]{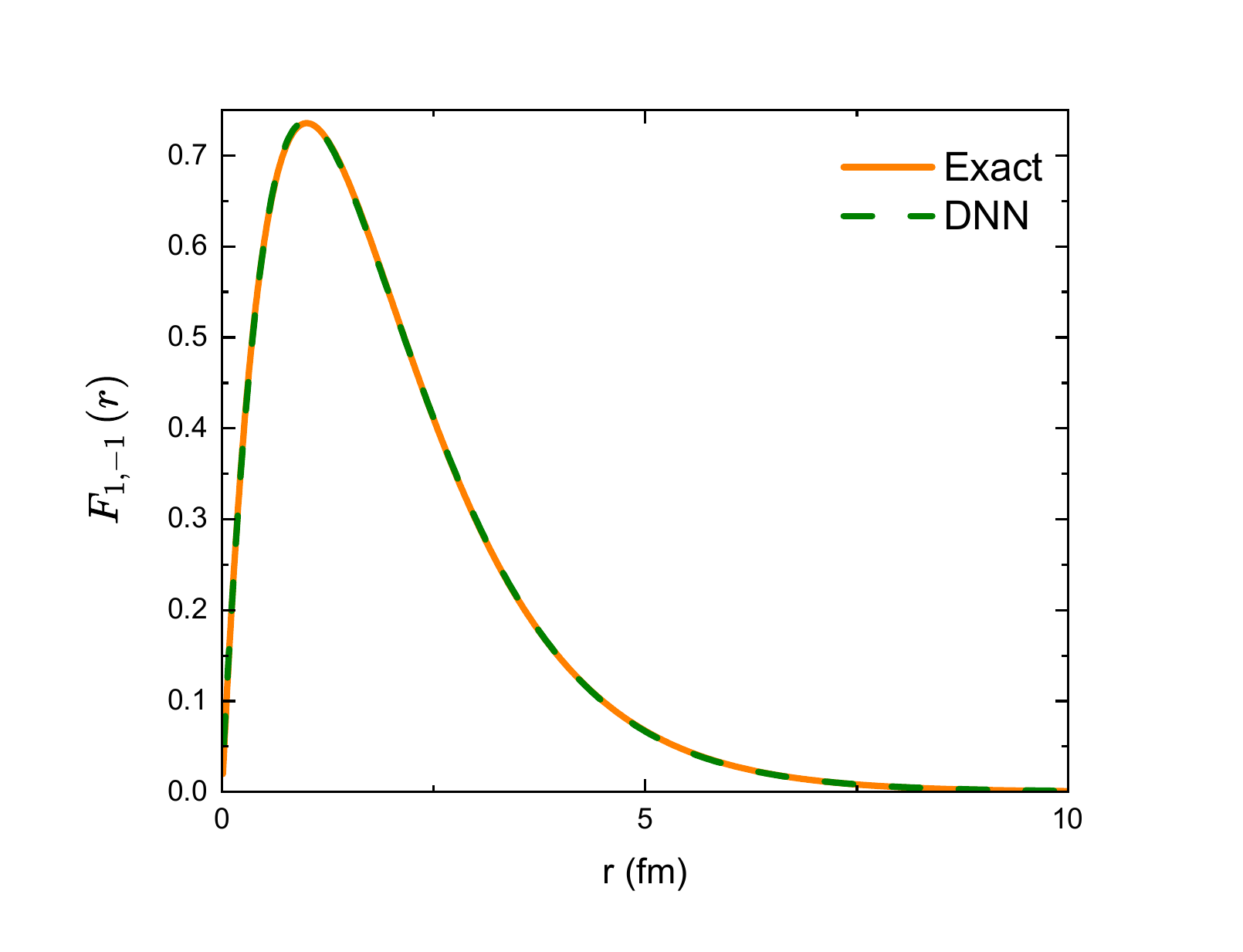} 
  \includegraphics[width=1.0\linewidth]{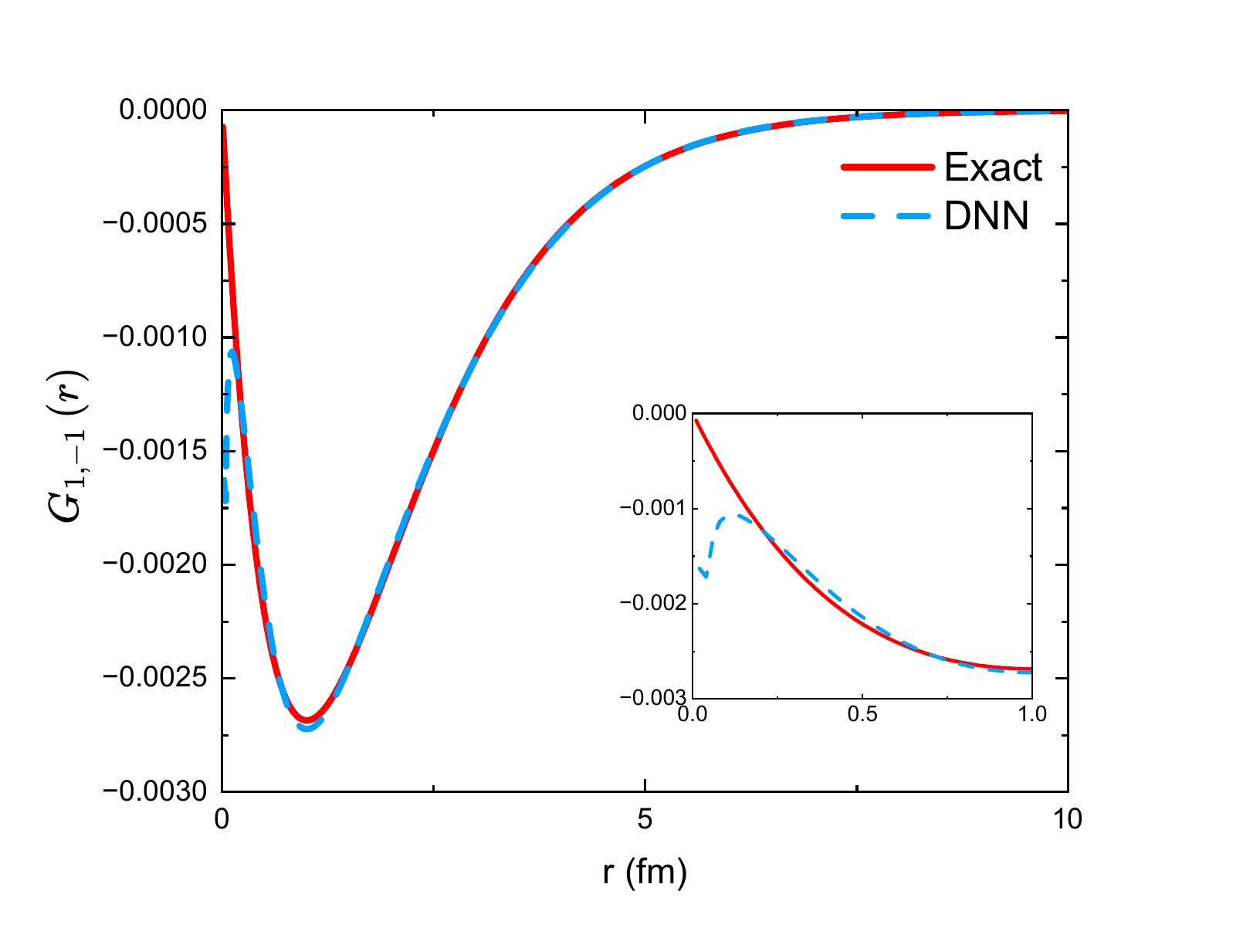}
  \caption{
    The $ F $- and $ G $-components of the ground-state wave functions of a hydrogen atom with using $ F_{n \kappa} $ as the DNN output.}
  \label{C_G_no_constrain}
\end{figure}
\begin{figure}[tb]
  \centering
  \includegraphics[width=1.0\linewidth]{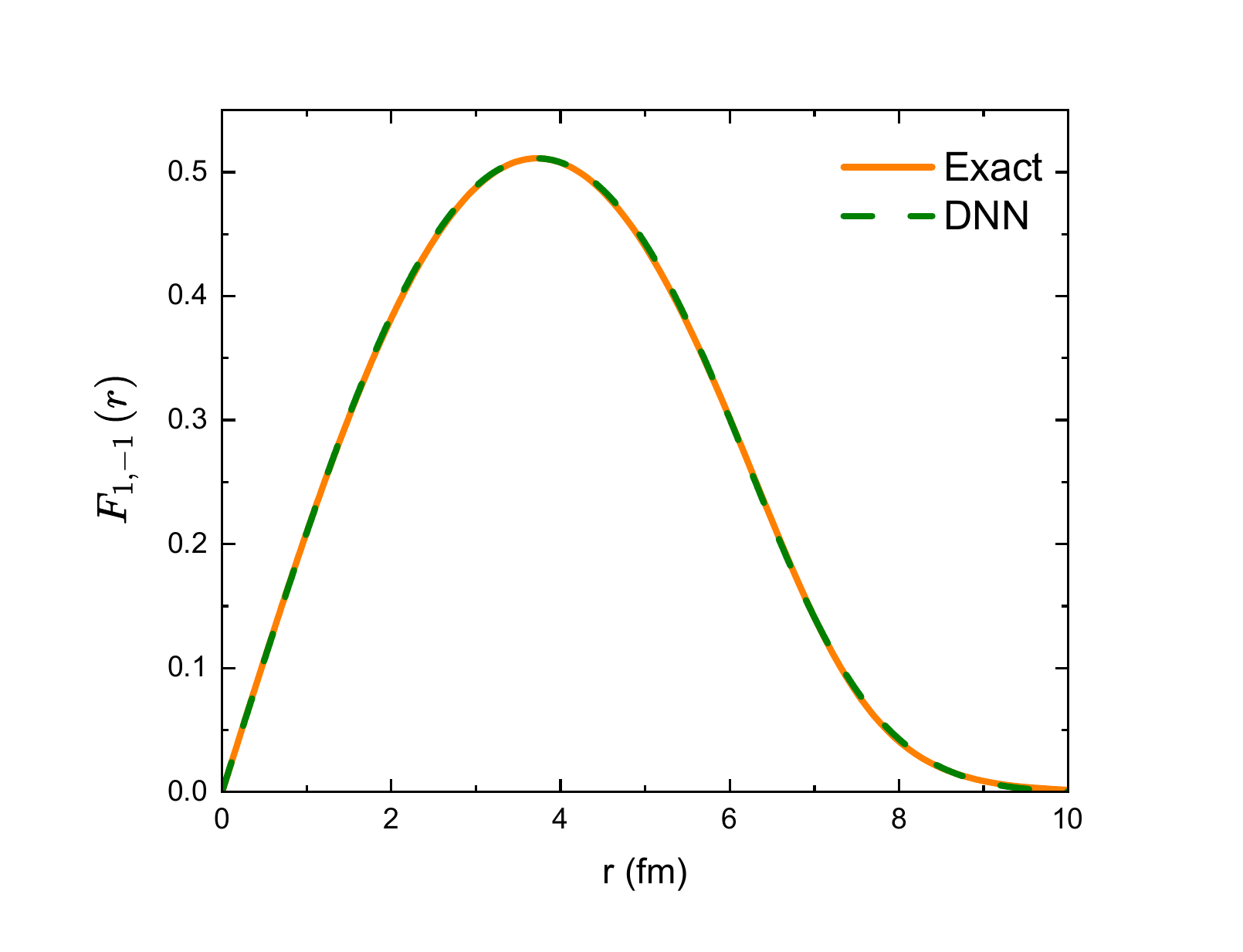} 
  \includegraphics[width=1.0\linewidth]{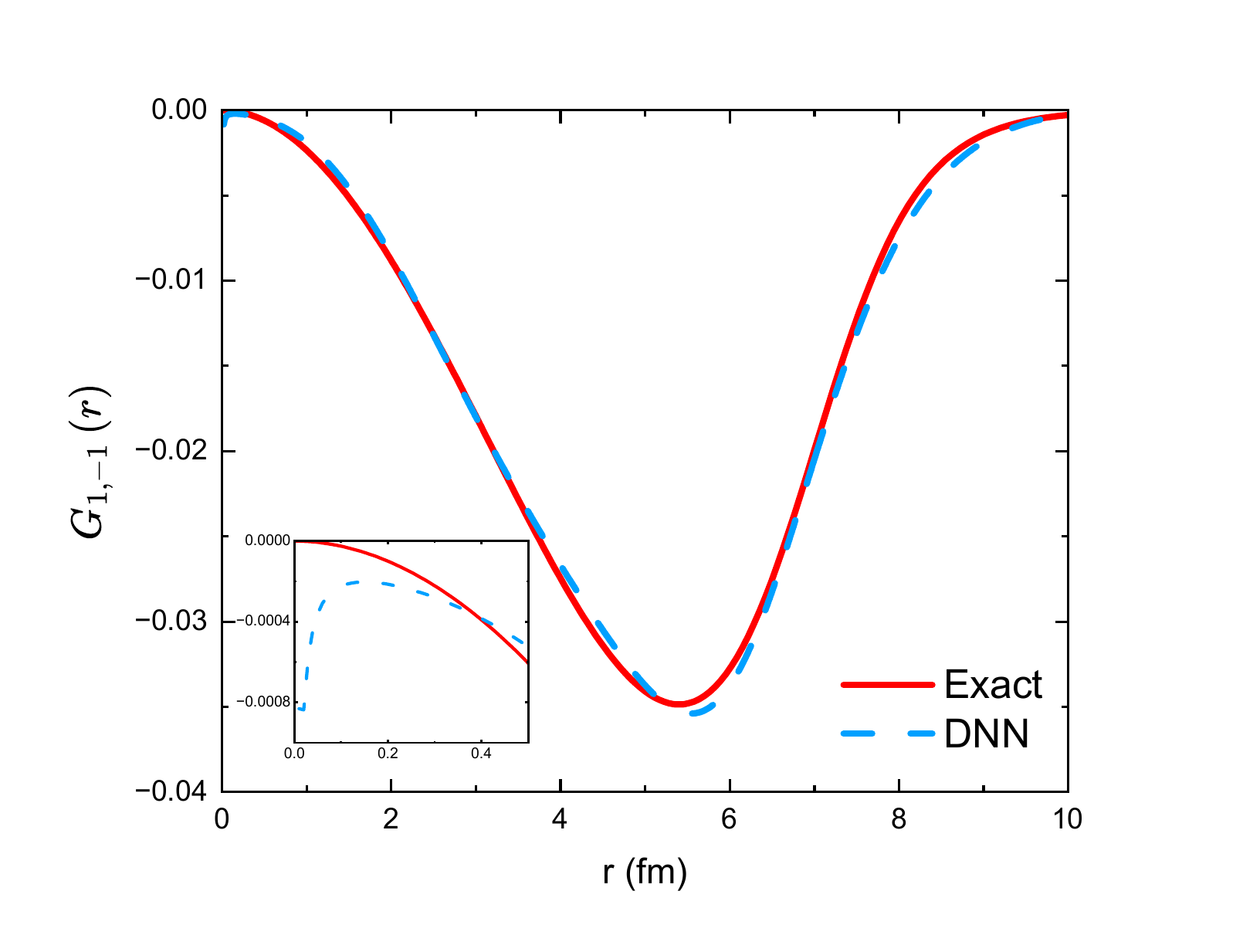}
  \caption{
    Same as Fig.~\ref{C_G_no_constrain} but for the Woods-Saxon potential for $ \nuc{Pb}{208}{} $.}
  \label{WS_G_no_constrain}
\end{figure}

If $ F_{n \kappa} $ is used as the DNN output, $ G_{n \kappa} $ of a hydrogen atom diverges at the origin as shown in Fig.~\ref{C_G_no_constrain}.
The ground-state wave functions of $ \nuc{Pb}{208}{} $ are shown in Fig.~\ref{WS_G_no_constrain}.
The diverge at the origin is smaller compared to the Coulomb case but still exists.
This may be because the gradient of $ F_{n \kappa} $ of $ \nuc{Pb}{208}{} $ at the origin is smaller compared with the hydrogen atom, which is easier for the DNN to optimize.
So that the $ G_{n \kappa} $ calculated by the $ F_{n \kappa} $ from Eq.~\eqref{eq9} is more accurate for $ \nuc{Pb}{208}{} $ than the hydrogen atom.
%
\subsection{Minimizing without the inverse Hamiltonian}
\label{Appendix:A}
\par 
Because of the existence of the Dirac sea, the ground state is no longer the lowest energy of the Dirac Hamiltonian $ H'_{\urm{D} r} $.
We employ the fully connected DNN in the main text and the non-fully connected DNN in Appendix~\ref{Appendix:B} to minimize the energy expectation value of $ H'_{\urm{D} r} $ 
\begin{equation}
  \epsilon_{\urm{D}}
  =
  \min
  \frac{\brakket{\varphi}{H'_{\urm{D} r}}{\varphi}}{\braket{\varphi}{\varphi}}
\end{equation}
as the loss function, where the Coulomb potential is adopted as a test case.
\par
\begin{figure}[tb]
  \centering
  \subfloat[Non-fully connected DNN]{\label{non-fully Dirac_sea}\includegraphics[width=1.0\linewidth]{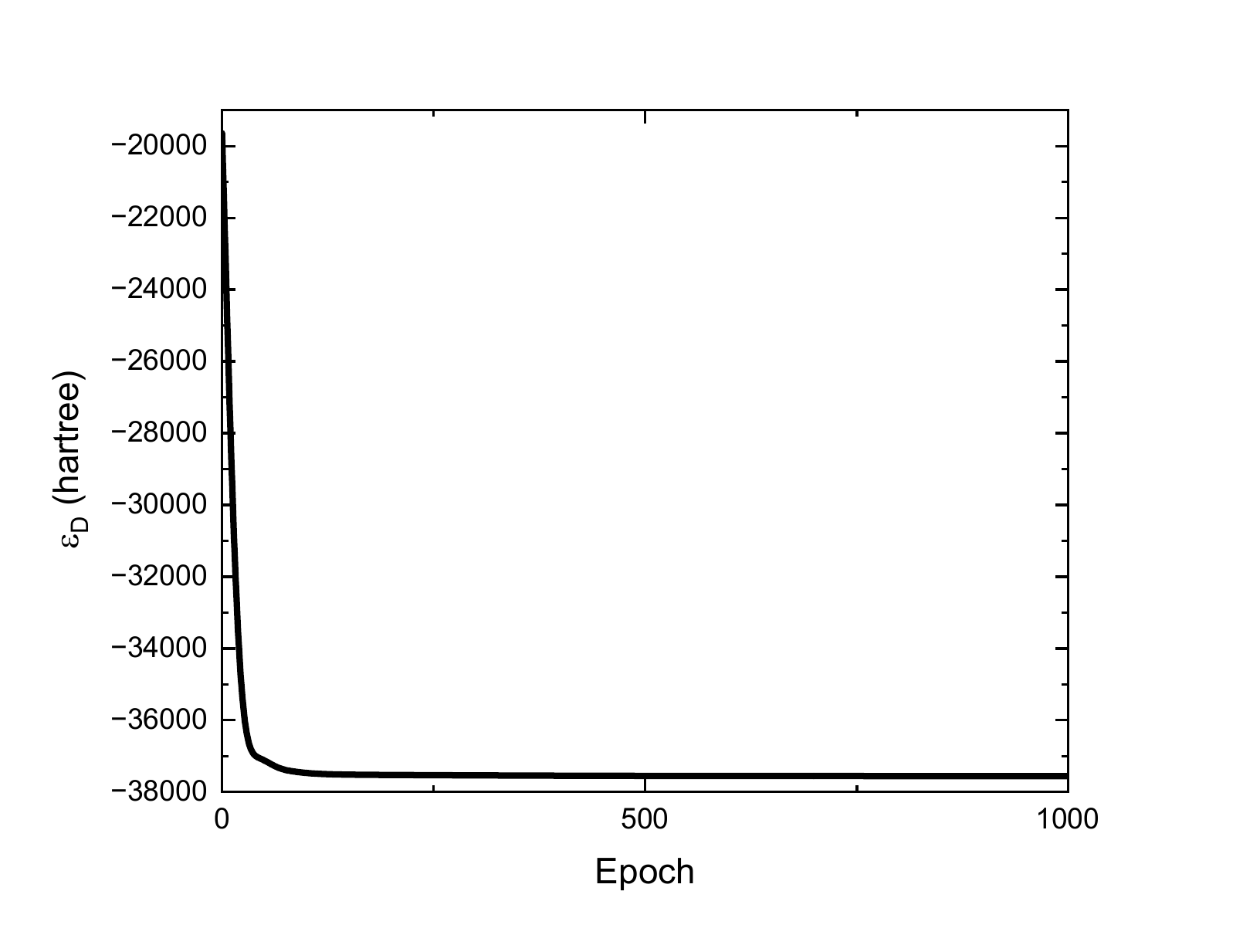}} 
  \quad
  \subfloat[Fully connected DNN]{\label{fully Dirac_sea}\includegraphics[width=1.0\linewidth]{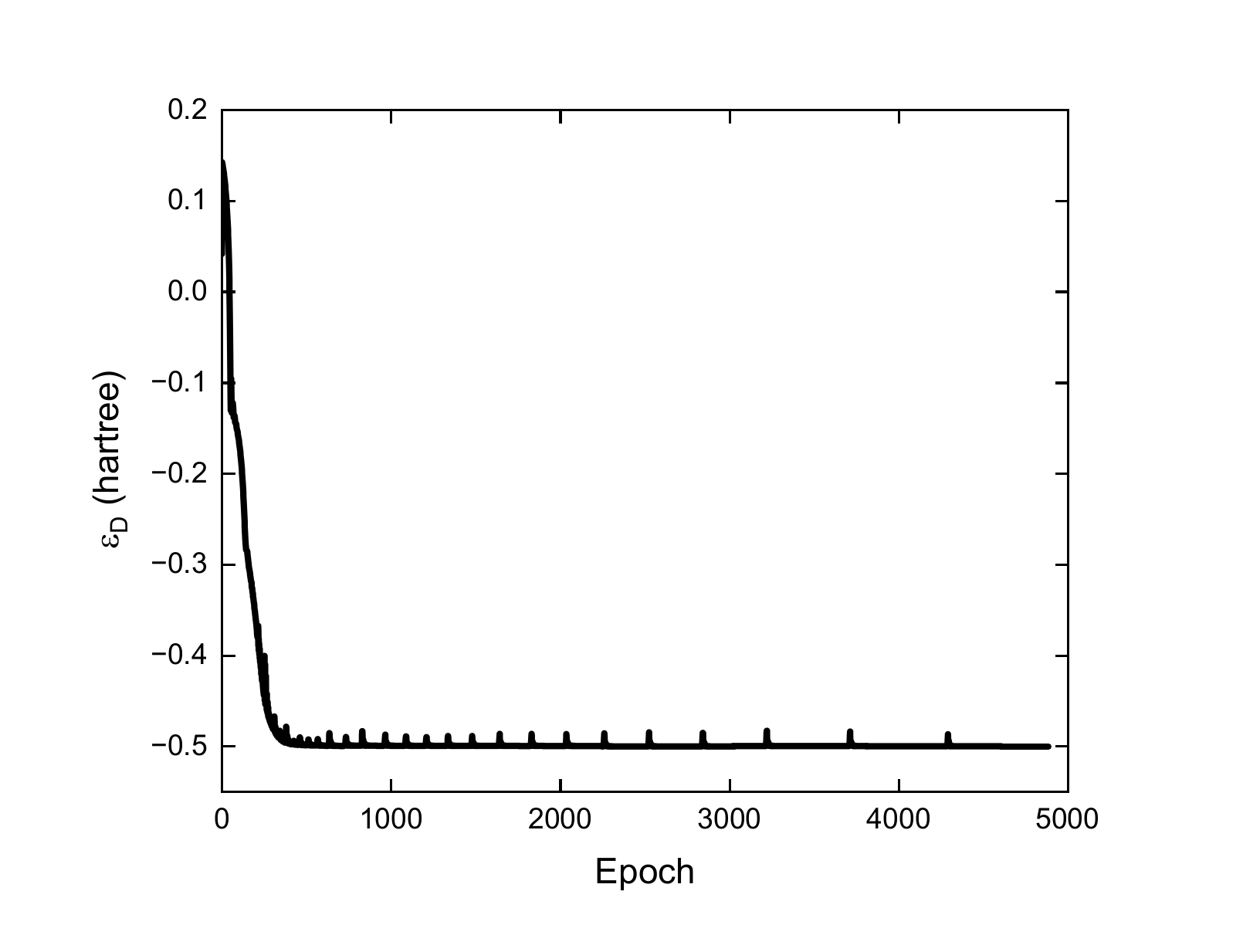}} 
  \caption{The energy of the Dirac Hamiltonian $ H'_{\urm{D} r} $ for a hydrogen atom with minimizing
    $ \epsilon_{\urm{D}} $ directly as a function of epochs.
    In the case of the non-fully connected DNN, the program doesn't converge and falls into the Dirac sea.
    In the case of the fully connected DNN, the loss function converges at the ground-state energy.}
  \label{Dirac_sea}
\end{figure}

As shown in Fig.~\ref{Dirac_sea}(a)
for the non-fully connected DNN, 
it is observed that $ \epsilon_{\urm{D}} $ falls into the Dirac sea, which is the so-called variational collapse.
For the fully connected DNN, as shown in Fig.~\ref{Dirac_sea}(b)
we discover that $ \epsilon_{\urm{D}} $ can converge at the ground-state energy.
This may because the local minimum of the ground state is too deep for the fully connected DNN so that the loss function is trapped.
\par
%
%
%
%
%
\end{document}